\documentclass[12pt]{iopart}
\usepackage{graphicx}
\usepackage{iopams}  
\usepackage{setstack}
\usepackage{amsfonts}
\usepackage{url}
\usepackage{color}
\usepackage{appendix}
\usepackage{cases}

\def\erfc{{\rm erfc}}
\def\erf{{\rm erf}}

\def\P{{\mathbb P}}

\def\L{\mathcal L}
\def\ve{\varepsilon}

\def\x{\hat{x}}

\def\D{{\mathcal D}}

\def\w{{w}}

\newcommand{\binom}[2]{{#1 \choose #2}}

\begin{document}

\title[First Exit Times of Harmonically Trapped Particles]
{First Exit Times of Harmonically Trapped Particles: \\ A Didactic Review}

\author{Denis~S.~Grebenkov}
  \ead{denis.grebenkov@polytechnique.edu}
\address{
$^1$ Laboratoire de Physique de la Mati\`{e}re Condens\'{e}e (UMR 7643), \\ 
CNRS -- Ecole Polytechnique, 91128 Palaiseau, France \\
$^2$ St. Petersburg National Research University of Information Technologies, \\
Mechanics and Optics, 197101 St. Petersburg, Russia  }

\date{\today}

\begin{abstract}
We revise the classical problem of characterizing first exit times of
a harmonically trapped particle whose motion is described by one- or
multi-dimensional Ornstein-Uhlenbeck process.  We start by recalling
the main derivation steps of a propagator using Langevin and
Fokker-Planck equations.  The mean exit time, the moment-generating
function, and the survival probability are then expressed through
confluent hypergeometric functions and thoroughly analyzed.  We also
present a rapidly converging series representation of confluent
hypergeometric functions that is particularly well suited for
numerical computation of eigenvalues and eigenfunctions of the
governing Fokker-Planck operator.  We discuss several applications of
first exit times such as detection of time intervals during which
motor proteins exert a constant force onto a tracer in optical
tweezers single-particle tracking experiments; adhesion bond
dissociation under mechanical stress; characterization of active
periods of trend following and mean-reverting strategies in
algorithmic trading on stock markets; relation to the distribution of
first crossing times of a moving boundary by Brownian motion.  Some
extensions are described, including diffusion under quadratic
double-well potential and anomalous diffusion.
\end{abstract}


%02.50.Ey 	Stochastic processes  (Probability theory, stochastic processes, and statistics)
%05.10.Gg 	Stochastic analysis methods (Fokker-Planck, Langevin, etc.) 
%05.40.-a 	Fluctuation phenomena, random processes, noise, and Brownian motion
%02.30.Gp 	Special functions

%\pacs{ 02.50.-r, 05.60.-k, 05.10.-a, 02.70.Rr }
% 02.50.-r  (Probability theory, stochastic processes, and statistics)
% 05.60.-k  (Transport processes)
% 05.10.-a  (Computational methods in statistical physics and nonlinear dynamics) 
% 02.70.Rr  (General statistical methods)


\submitto{\JPA}

\maketitle
\tableofcontents

\section{Introduction}

First passage time (FPT) distributions have found numerous
applications in applied mathematics, physics, biology and finance
\cite{Redner,Metzler,Benichou14}.  The FPT can characterize the time
needed for an animal to find food; the time for an enzyme to localize
specific DNA sequence and to initiate biochemical reaction; the time
to exit from a confining domain (e.g., a maze); or the time to buy or
sell an asset when its price deviation from the mean exceeds a
prescribed threshold.  The FPT distribution has been studied for a
variety of diffusive processes, ranging from ordinary diffusion
(Brownian motion) to continuous-time random walks (CTRWs)
\cite{Bouchaud90,Metzler00,Metzler04,Condamin07b,Yuste07,Yuste08,Grebenkov10a,Grebenkov10b},
fractional Brownian motion
\cite{Molchan99,Likthman06,Jeon11b,Sanders12}, L\'evy flights
\cite{Koren07a,Koren07b,Tejedor11}, surface-mediated diffusion
\cite{Benichou10,Benichou11,Rupprecht12a,Rupprecht12b} and other
intermittent processes \cite{Oshanin09,Benichou11a}, diffusion in
scale invariant media \cite{Condamin07,Condamin08}, trapped diffusion
\cite{Holcman05}, thermally driven oscillators \cite{Crandall70},
Ornstein-Uhlenbeck process
\cite{Siegert51,Darling53,Lindenberg75,Ricciardi88,Leblanc00,Going03,Alili05,Yi10,Spendier13,Toenjes13},
and many others
\cite{Redner,Metzler,Borodin,Revuz,Ito,Jeanblanc,Crank,Carslaw,Hughes,Gardiner,Doob,vanKampen}.

In this review, we revise the classical problem of characterizing the
first {\it exit} time (FET) distribution of a multi-dimensional
Ornstein-Uhlenbeck process from a ball
\cite{Uhlenbeck30,Wang45,Ricciardi}.  The probability distribution can
be found through the inverse Fourier (resp. Laplace) transform of the
characteristic (resp. moment-generating) function for which explicit
representations in terms of special functions are well known
\cite{Darling53,Borodin}.  Although the problem is formally solved,
the solution involves confluent hypergeometric functions and thus
requires subtle asymptotic methods and computational hints.  The aim
of the review is to provide a didactic self-consistent description of
theoretical, numerical and practical aspects of this problem.

First, we recall the main derivation steps of the FET distribution,
from the Langevin equation (Sec. \ref{sec:Langevin}), through
forward and backward Fokker-Planck (FP) equations
(Sec. \ref{sec:forward_FP}, \ref{sec:backward_FP}), to spectral
decompositions based on the eigenvalues and eigenfunctions of the FP
operator (Sec. \ref{sec:first}).  This general formalism is then
applied to describe the first exit times of harmonically trapped
particles in one dimension: the mean exit time (Sec. \ref{sec:mean}),
the survival probability (Sec. \ref{sec:spectral_1d}), and the
moment-generating function (Sec. \ref{sec:generating_1d}).  In
particular, we analyze the asymptotic behavior of the mean exit time
and eigenvalues in different limits (e.g., strong trapping potential,
large constant force, etc.).  Extensions to the radial
Ornstein-Uhlenbeck process in higher-dimensional cases for both
interior and exterior problems are presented in Sec. \ref{sec:higher}
and Sec. \ref{sec:exterior}, respectively.  Although most of these
results are classical, their systematic self-contained presentation
and numerical illustrations are missing.

Section \ref{sec:discussion} starts from the summary of computational
hints for computing confluent hypergeometric functions while technical
details are reported in \ref{sec:Kummer}.  We discuss then three
applications: (i) calibration of optical tweezers' stiffness in
single-particle tracking experiments and detection of eventual
constant forces exerted on a tracer by motor proteins
(Sec. \ref{sec:tracking}), (ii) adhesion bond dissociation under
mechanical stress (Sec. \ref{sec:bond}), and (iii) distribution of
triggering times of trend following strategies in algorithmic trading
on stock markets (Sec. \ref{sec:trading}).  We also illustrate a
direct relation to the distribution of first crossing times of a
moving boundary by Brownian motion (Sec. \ref{sec:crossing}).
Finally, we present several extensions of the spectral approach,
including diffusion under quadratic double-well potential
(Sec. \ref{sec:double-well}) and anomalous diffusion
(Sec. \ref{sec:extensions}).  Many technical details are summarized in
Appendices.

\section{First exit time distribution}
\label{sec:theory}

We first recall the standard theoretical description of harmonically
trapped particles by Langevin and Fokker-Planck equations
\cite{Gardiner,Coffey,Risken}.  We start with one-dimensional
Ornstein-Uhlenbeck process and then discuss straightforward extensions
to higher dimensions.

\subsection{Langevin equation}
\label{sec:Langevin}

We consider a diffusing particle of mass $m$ trapped by a harmonic
potential of strength $k$ and pulled by a constant force $F_0$.  The
thermal bath surrounding the particle results in its stochastic
trajectory which can be described by Langevin equation \cite{Coffey}
\begin{equation}
m \ddot{X}(t) = - \gamma \dot{X}(t) + F(X(t)) + \xi(t) ,
\end{equation}
where $-\gamma \dot{X}(t)$ is the viscous Stokes force ($\gamma$ being
the drag constant), $F(X(t)) = - kX(t) + F_0$ includes the externally
applied Hookean and constant forces, and $\xi(t)$ is the thermal
driving force with Gaussian distribution such that $\langle \xi(t)
\rangle = 0$ and $\langle \xi(t) \xi(t')\rangle = 2k_B T\gamma
\delta(t-t')$, with $k_B \simeq 1.38 \cdot 10^{-23}~ J/K$ being the
Boltzmann constant, $T$ the absolute temperature (in degrees Kelvin),
$\delta(t)$ the Dirac distribution, and $\langle \ldots \rangle$
denoting the ensemble average or expectation.  In the overdamped limit
($m = 0$), one gets the first-order stochastic differential equation
\begin{equation}
\label{eq:Langevin2}
\dot{X}(t) =  \frac{1}{\gamma} \bigl[F(X(t)) + \xi(t)\bigr] = \frac{k}{\gamma} (\x - X(t)) + \frac{\xi(t)}{\gamma}  ,  \qquad X(0) = x_0,
\end{equation}
where $\x = F_0/k$ is the stationary position (mean value), and $x_0$
is the starting position.  The Langevin equation can also be written
in a conventional (dimensionless) stochastic form \cite{Borodin,Revuz}
\begin{equation}
\label{eq:Langevin2a}
dX_t = \mu(X_t,t)dt + \sigma(X_t,t) dW_t ,  \qquad X_0 = x_0,
\end{equation}
where $W_t$ is the standard Wiener process (Brownian motion),
$\mu(x,t)$ and $\sigma(x,t)$ are the drift and volatility which in
general can depend on $x$ and $t$.  In our case, the volatility is
constant, while the drift is a linear function of $x$, $\mu(x,t) = (\x
- x)\theta$, i.e.
\begin{equation}
\label{eq:Langevin3}
dX_t = \theta(\x - X_t)dt + \sigma dW_t ,  \qquad X_0 = x_0,
\end{equation}
where $\theta = k\delta /\gamma$, and $\sigma = \sqrt{2D\delta}$, with
$\delta$ being a time scale, and $D = k_B T/\gamma$ the diffusion
coefficient.  This stochastic differential equation defines an
Ornstein-Uhlenbeck (OU) process, with mean $\x$, variance $\sigma^2$,
and rate $\theta$.  An integral representation of
Eq. (\ref{eq:Langevin3}) reads
\begin{equation}
X_t = x_0 e^{-\theta t} + \x(1 - e^{-\theta t}) + \sigma \int\limits_0^t e^{\theta(t'-t)} dW_{t'} .
\end{equation}
One can see that $X_t$ is a Gaussian process with mean $\langle
X_t\rangle = x_0 e^{-\theta t} + \x(1-e^{-\theta t})$ and covariance
$\langle X_t X_{t'} \rangle - \langle X_t\rangle \langle X_{t'} \rangle 
= \frac{\sigma^2}{2\theta} (e^{-\theta|t-t'|} - e^{-\theta(t+t')})$.

The discrete version of Eq. (\ref{eq:Langevin3}) with a fixed time
step $\delta$ is known as auto-regressive model AR(1):
\begin{equation}
\label{eq:AR}
X_n = (1 - k\delta/\gamma) X_{n-1} + F_0 \delta/\gamma + \sqrt{2D\delta}~ \xi_n,
\end{equation}
where $\xi_n$ are standard iid Gaussian variables with mean zero and
unit variance.  This discrete scheme can be used for numerical
generation of stochastic trajectories.  An extension of the above
stochastic description to multi-dimensional processes is
straightforward.

\subsection{Forward Fokker-Planck equation}
\label{sec:forward_FP}

The Langevin equation (\ref{eq:Langevin2}) expresses the displacement
$\dot{X}(t) \delta$ over a short time step $\delta$ in terms of the
current position $X(t)$.  In other words, the distribution of the next
position is fully determined by the current position, the so-called
Markov property.  Such a Markov process can be characterized by a
propagator or a transition density, i.e., the conditional probability
density $p(x,t|x_0,t_0)$ of finding the particle at $x$ at time $t$,
given that it was at $x_0$ at earlier time $t_0$.  The propagator can
be seen as a ``fraction'' of paths from $x_0$ to $x$ among all paths
started at $x_0$ (of duration $t-t_0$) which formally writes as the
average of the Dirac distribution $\delta(X(t)-x)$ over all random
paths started from $x_0$: $p(x,t|x_0,t_0) = \langle \delta(X(t)-x)
\rangle_{X(t_0)=x_0}$.  The Markov property implies the
Chapman-Kolmogorov (or Smoluchowski) equation
\begin{equation}
\label{eq:CK}
p(x,t|x_0,t_0) = \int\limits_{-\infty}^\infty dx' ~ p(x,t|x',t') ~ p(x',t'|x_0,t_0) \qquad (t_0 < t' < t),
\end{equation}
which expresses a simple fact that any continuous path from $X(t_0) =
x_0$ to $X(t) = x$ can be split at any intermediate time $t'$ into two
{\it independent} paths, from $x_0$ to $x'$, and from $x'$ to $x$.

As a function of the arrival state ($x$ and $t$), the propagator
satisfies the forward Fokker-Planck (FP) equation
\cite{Coffey,Risken}.  We reproduce the derivation of this equation
from \cite{Kolpas07} which relies on the evaluation of the integral
\begin{equation*}
I = \int\limits_{-\infty}^\infty dx ~ h(x) \biggl[p(x,t+\delta|x_0,t_0) - p(x,t|x_0,t_0)\biggr]
\end{equation*}
for any smooth function $h(x)$ with compact support.  One has
\begin{eqnarray*}
I &=& \int\limits_{-\infty}^\infty dx ~ h(x) \int\limits_{-\infty}^\infty dx'~ p(x,t+\delta|x',t)~ p(x',t|x_0,t_0) \\
&-& \int\limits_{-\infty}^\infty dx' ~ h(x')~ p(x',t|x_0,t_0)  \int\limits_{-\infty}^\infty dx~  p(x,t+\delta|x',t) \\
&=& \int\limits_{-\infty}^\infty dx \int\limits_{-\infty}^\infty dx'~ p(x,t+\delta|x',t)~ p(x',t|x_0,t_0)~ \biggl[h(x) - h(x')\biggr],
\end{eqnarray*}
where the first term was represented using Eq. (\ref{eq:CK}), while
the normalization of the probability density $p(x,t+\delta|x',t)$
allowed one to add the integral over $x$ in the second term.
Expanding $h(x)$ into a Taylor series around $x'$ and then exchanging
the integration variables $x$ and $x'$, one gets
\begin{equation*}
\fl
I = \int\limits_{-\infty}^\infty dx ~p(x,t|x_0,t_0)~ \sum\limits_{n=1}^{\infty} 
\left(\frac{d^n}{dx^n} h(x)\right) ~ \frac{1}{n!} \int\limits_{-\infty}^\infty dx' ~p(x',t+\delta|x,t) ~ (x' - x)^n ,
\end{equation*}
Finally, integrating each term by parts $n$ times, dividing by
$\delta$ and taking the limit $\delta\to 0$ yield
\begin{equation*}
\fl
\int\limits_{-\infty}^\infty dx ~ h(x)~ \frac{\partial p(x,t|x_0,t_0)}{\partial t} = 
\int\limits_{-\infty}^\infty dx ~ h(x) \sum\limits_{n=1}^{\infty} (-1)^n \frac{d^n}{dx^n} \biggl(D^{(n)}(x) ~ p(x,t|x_0,t_0)\biggr) ,
\end{equation*}
where the left hand side is the limit of $I/\delta$ as $\delta\to 0$,
and
\begin{equation}
\label{eq:Dn}
D^{(n)}(x) = \frac{1}{n!} \lim\limits_{\delta\to 0} \frac{1}{\delta} \int\limits_{-\infty}^\infty dx' ~p(x',t+\delta|x,t) ~ (x' - x)^n .
\end{equation}
Since the above integral relation is satisfied for arbitrary function
$h(x)$, one deduces the so-called Kramers-Moyal expansion:
\begin{equation}
\frac{\partial p(x,t|x_0,t_0)}{\partial t} = \sum\limits_{n=1}^{\infty} (-1)^n \frac{d^n}{dx^n} \biggl(D^{(n)}(x) ~ p(x,t|x_0,t_0)\biggr) .
\end{equation}
Here we assumed that the process is time homogeneous, i.e.,
$p(x,t|x_0,t_0)$ is invariant under time shift: $p(x,t|x_0,t_0) =
p(x,t+t'|x_0,t_0+t')$ that implies the time-independence of
$D^{(n)}(x)$.

The density $p(x',t+\delta|x,t)$ in Eq. (\ref{eq:Dn}) characterizes
the displacement between $X(t) = x$ and $X(t+\delta) = x'$ which can
be written as $X(t+\delta)-X(t) \simeq \frac{\delta}{\gamma} [F(x) +
\xi(t)]$ for small $\delta$ according to the Langevin equation
(\ref{eq:Langevin2}).  After discretization in units of $\delta$, the
thermal force $\xi(t)$ becomes a Gaussian variable with mean zero and
variance $2k_B T \gamma/\delta$.  As a consequence, the displacement
$x'-x$ is also a Gaussian variable with mean $(\delta/\gamma) F(x)$
and variance $(\delta/\gamma)^2 ~ 2k_B T\gamma/\delta$, i.e.,
\begin{equation*}
p(x',t+\delta|x,t) = \frac{1}{\sqrt{4\pi D \delta}} \exp\left(- \frac{(x' - x - F(x) \delta/\gamma)^2}{4D\delta} \right)
\end{equation*}
for small $\delta$.  Substituting this density into Eq. (\ref{eq:Dn})
and evaluating Gaussian integrals, one gets $D^{(1)}(x) =
F(x)/\gamma$, $D^{(2)} = D$, and $D^{(n)} = 0$ for $n > 2$ that yields
the forward Fokker-Planck equation
\begin{equation}
\label{eq:FP_operator0}
\frac{\partial}{\partial t}~ p(x,t|x_0,t_0) = \L_x~  p(x,t|x_0,t_0) ,  \qquad \L_x = - \partial_x \frac{F(x)}{\gamma} + D\partial_x^2 ,
\end{equation}
where $\L_x$ is the Fokker-Planck operator acting on the arrival point
$x$.  This equation is completed by the initial condition
$p(x,t_0|x_0,t_0) = \delta(x-x_0)$ at $t = t_0$, with a fixed starting
point $x_0$.  Note that the forward FP equation can be seen as the
probability conservation law,
\begin{equation*}
\frac{\partial}{\partial t}~ p(x,t|x_0,t_0) = - \partial_x J(x,t|x_0,t_0), 
\end{equation*}
where $J(x,t|x_0,t_0) = \frac{F(x)}{\gamma} p(x,t|x_0,t_0) -
D\partial_x p(x,t|x_0,t_0)$ is the probability flux.  Setting $J = 0$,
one can solves the first-order differential equation to retrieve the
equilibrium solution $p_{\rm eq}(x) = Z \w(x)$, where $Z$ is the
normalization factor, and
\begin{equation}
\label{eq:weight}
\fl
\w(x) = \exp\left(\int\limits_0^x dx' \frac{F(x')}{k_B T}\right) = \exp\left(- \frac{V(x)}{k_B T}\right) 
= \exp\left(-\frac{kx^2}{2k_B T} + \frac{F_0 x}{k_B T}\right) ,
\end{equation}
where $V(x) = - \int_0^x dx' F(x')$ is the potential associated to the
force $F(x)$.  This is the standard Boltzmann-Gibbs equilibrium
distribution.

When the FP operator $\L_x$ has a discrete spectrum, the probability
density admits the spectral decomposition
\begin{equation}
\label{eq:p_spectral0}
p(x,t|x_0,t_0) = \sum\limits_{n=0}^\infty  v_n(x)~ v_n(x_0)~ \tilde{\w}(x_0)~ e^{-\lambda_n (t-t_0)}  
\end{equation}
over the eigenvalues $\lambda_n$ and eigenfunctions $v_n(x)$ of
$\L_x$:
\begin{equation}
\label{eq:eigen}
\L_x v_n(x) + \lambda_n v_n(x) = 0 \qquad (n = 0,1,2,\ldots)
\end{equation}
(eventually with appropriate boundary conditions, see below).  The
weight $\tilde{\w}(x) = 1/\w(x)$ ensures the orthogonality of
eigenfunctions:
\begin{equation}
\int\limits dx ~ \tilde{\w}(x) ~ v_m(x)~ v_n(x) = \delta_{m,n} ,
\end{equation}
while the closure (or completeness) relation reads
\begin{equation}
\sum\limits_{n=0}^\infty  v_n(x)~ v_n(x_0) ~ \tilde{\w}(x_0) = \delta(x-x_0) .
\end{equation}
This relation implies the initial condition $p(x,t_0|x_0,t_0) =
\delta(x-x_0)$.  As for the Langevin equation, an extension to the
multi-dimensional case is straightforward.  In particular, the
derivative $\partial_x$ is replaced by the gradient operator, while
$\partial_x^2$ becomes the Laplace operator
\cite{Risken,Grebenkov13}.

\subsection{Backward Fokker-Planck equation}
\label{sec:backward_FP}

The forward FP equation describes the evolution of the probability
density $p(x,t|x_0,t_0)$ from a given initial state (here, the
starting point $x_0$ at time $t_0$).  Alternatively, if the particle
is found at the arrival point $x$ at time $t$ (or, more generally, in
a prescribed subset of states), one can interpret $p(x,t|x_0,t_0)$ as
the conditional probability density that the particle is started from
$x_0$ at time $t_0$ knowing that it arrived at $x$ at later time $t$.
As a function of $x_0$ and $t_0$, this probability density satisfies
the {\it backward} Fokker-Planck (or Kolmogorov) equation
\cite{Risken}:
\begin{equation}
\label{eq:BFP}
- \frac{\partial}{\partial t_0}~ p(x,t|x_0,t_0) = \L_{x_0}^*
~p(x,t|x_0,t_0),
\end{equation}
where the backward FP operator $\L^*$ is adjoint to the forward FP
operator $\L$ (i.e., $(\L f, g) = (f, \L^* g)$ for any two functions
$f$ and $g$ from an appropriate functional space).
Eq. (\ref{eq:FP_operator0}) implies
\begin{equation}
\label{eq:FP_operator}
\L^*_{x_0} = \frac{F(x_0)}{\gamma} \partial_{x_0} + D \partial^2_{x_0} = \frac{k}{\gamma}(\x - x_0) \partial_{x_0} + D \partial^2_{x_0} .
\end{equation}
Note that this operator acts on the starting point $x_0$ while the
sign minus in front of time derivative reflects the backward time
direction.  Eq. (\ref{eq:BFP}) is easily obtained by differentiating
the Champan-Kolmogorov equation (\ref{eq:CK}) with respect to the
intermediate time $t'$.

The eigenvalues of both forward and backward FP operators are
identical, while the eigenfunctions $u_n(x)$ of the backward FP
operator $\L^*$ are simply $u_n(x) = v_n(x)/\w(x)$.  As a consequence,
one can rewrite the spectral decomposition (\ref{eq:p_spectral0}) as
\begin{equation}
\label{eq:p_spectral}
p(x,t|x_0,t_0) = \sum\limits_{n=0}^\infty  u_n(x_0)~ u_n(x)~ \w(x)~ e^{-\lambda_n (t-t_0)}  ,
\end{equation}
with the weight $\w(x)$ from Eq. (\ref{eq:weight}).  The
eigenfunctions $u_n(x)$ are as well orthogonal:
\begin{equation}
\int\limits dx ~ \w(x) ~ u_m(x)~ u_n(x) = \delta_{m,n} ,
\end{equation}
while the closure (or completeness) relation reads
\begin{equation}
\sum\limits_{n=0}^\infty  u_n(x_0) ~u_n(x)~ \w(x) = \delta(x-x_0) .
\end{equation}
This relation implies the terminal condition $p(x,t|x_0,t) =
\delta(x-x_0)$ at $t_0 = t$.  In contrast to
Eq. (\ref{eq:p_spectral0}), the weight $\w(x)$ in the spectral
representation (\ref{eq:p_spectral}) depends on the {\it fixed}
arrival point $x$, while the backward FP operator $\L_{x_0}^*$ acts on
eigenfunctions $u_n(x_0)$.

When there is no force term, the operator $\L$ is self-adjoint, $\L =
\L^*$, and the probability density is invariant under time reversal:
$p(x,t|x_0,t_0) = p(x_0,t_0|x,t)$.  This property does not hold in the
presence of force.

Finally, the backward FP equation is closely related to the
Feynman-Kac formula for determining distributions of various Wiener
functionals \cite{Kac49,Kac51,Freidlin,Simon,Bass}.  For instance, we
already mentioned that the probability density $p(x,t|x_0,t_0)$ can be
understood as the conditional expectation: $p(x,t|x_0,t_0) = \langle
\delta(X(t) - x)\rangle_{X(t_0) = x_0}$.  More generally, for given
functions $\psi(x_0)$, $f(x_0,t_0)$ and $U(x_0,t_0)$, the conditional
expectation
\begin{eqnarray}
u(x_0,t_0) &=& \Biggl\langle \exp\Biggl(-\int\limits_{t_0}^{t} dt' U(X(t'),t')\Biggr) \psi(X(t)) \\
\nonumber
&+& \int\limits_{t_0}^t dt' f(X(t'),t') \exp\Biggl(-\int\limits_{t_0}^{t'} dt'' U(X(t''),t'')\Biggr) \Biggr\rangle_{X(t_0)=x_0}
\end{eqnarray}
satisfies the backward FP equation
\begin{equation}
- \frac{\partial}{\partial t_0} u(x_0,t_0) = \L^*_{x_0} u(x_0,t_0) - U(x_0,t_0) u(x_0,t_0) + f(x_0,t_0) ,
\end{equation}
subject to the terminal condition $u(x_0,t) = \psi(x_0)$ at a later
time $t > t_0$.

\subsection{First exit times}
\label{sec:first}

In this review, we study the random variable $\tau = \inf\{ t>0 ~:~
|X(t)| > L\}$, i.e., the first exit time of the process $X(t)$ from an
interval $[-L,L]$ when started from $x_0$ at $t_0 = 0$.  The
cumulative distribution function of $\tau$ is related to the
survival probability $S(x_0,t) = \P\{ \tau > t\}$ up to time $t$ of a
particle which started from $x_0$.  The notion of {\it survival} is
associated to disappearing of the particle that hit either endpoint,
due to chemical reaction, permeation, adsorption, relaxation,
annihilation, transformation or any other ``killing'' mechanism.  The
survival probability $S(x_0,t)$ can be expressed through the
probability density $p(x,t|x_0,0)$ of moving from $x_0$ to $x$ in time
$t$ {\it without visiting the endpoints $\pm L$ during this motion}.
Alternatively, $p(x,t|x_0,0)$ can be seen as the conditional
probability density of starting from point $x_0$ at time $t_0 = 0$
under condition to be at $x$ at time $t$.  This condition includes the
survival up to time $t$, i.e., not visiting the endpoints $\pm L$.
The probability density $p(x,t|x_0,0)$ satisfies the backward FP
equation with Dirichlet boundary condition at $x_0 = \pm L$:
$p(x,t|\pm L,0) = 0$.  This condition simply states that a particle
started from either endpoint has immediately hit this endpoint,
i.e. not survived.  Note that this condition is preserved during all
intermediate times $t'$ due to the Chapman-Kolmogorov equation
(\ref{eq:CK}).

Since the survival probability $S(x_0,t)$ ignores the actual position
$x$ at time $t$, one just needs to average the density $p(x,t|x_0,0)$
over $x$:
\begin{equation}
\label{eq:S_spectral}
S(x_0,t) = \int\limits_{-L}^L dx ~ p(x,t|x_0,0) 
 = \sum\limits_{n=0}^\infty  u_n(x_0) ~ e^{-\lambda_n t} \int\limits_{-L}^L dx~ u_n(x) \w(x) ,
\end{equation}
where the spectral decomposition (\ref{eq:p_spectral}) was used.  The
eigenfunctions $u_n(x)$ of the backward FP operator should satisfy
Dirichlet boundary condition at $x_0 = \pm L$: $u_n(\pm L) = 0$.
Eq. (\ref{eq:S_spectral}) also implies the backward FP equation
\begin{equation}
\label{eq:FP}
\frac{\partial S(x_0,t)}{\partial t} = \L^*_{x_0} S(x_0,t) ,
\end{equation}
which is completed by the initial condition $S(x_0,0) = 1$ (the
particle exists at the beginning) and Dirichlet boundary condition
$S(\pm L, t) = 0$ (the process is stopped upon the first arrival at
either endpoint of the confining interval $[-L,L]$).  Since the
process is homogeneous in time, $p(x,t|x_0,t_0)$ depends on $t-t_0$
and thus $\frac{\partial}{\partial t_0} p(x,t|x_0,t_0) = -
\frac{\partial}{\partial t} p(x,t|x_0,t_0)$ that allows one to write
the left-hand side of the backward FP equation (\ref{eq:FP}) with the
plus sign.  Note that the characterization of first passage times
through the backward FP equation goes back to the seminal work in 1933
by Pontryagin {\it et al.} \cite{Pontryagin33}.  Similar equations
emerge in quantum mechanics when one searches for eigenstates of a
particle trapped by a short-range harmonic potential \cite{Castro13}
(see also \ref{sec:QH} for quantum harmonic oscillator).

The FET probability density is $q(x_0,t) = - \frac{\partial
S(x_0,t)}{\partial t}$, while the moment-generating function is given
by its Laplace transform:
\begin{equation}
\langle e^{-s \tau}\rangle = \int\limits_0^\infty dt~ e^{-st} q(x_0,t) \equiv \tilde{q}(x_0,s) ,
\end{equation}
with tilde denoting Laplace-transformed quantities.  The Laplace
transform of Eq. (\ref{eq:FP}) yields the equation $(\L^*_{x_0} -
s)\tilde{S}(x_0,s) = -1$ with Dirichlet boundary conditions.  Since
$\tilde{q}(x_0,s) = 1 - s\tilde{S}(x_0,s)$, one gets
\begin{equation}
\label{eq:FP_char}
\bigl(\L^*_{x_0} - s \bigl)\tilde{q}(x_0,s) = 0,
\end{equation}
with Dirichlet boundary condition $\tilde{q}(\pm L,s) = 1$.

Finally, the moments $\langle \tau^m \rangle_{x_0}$ can be found in
one of standard ways:
\begin{enumerate}
\item
from the moment-generating function,
\begin{equation}
\langle \tau^m \rangle_{x_0} = (-1)^m \lim\limits_{s\to 0} \frac{\partial^m}{\partial s^m}~ \tilde{q}(x_0,s) ;
\end{equation}

\item 
from the spectral representation of the survival probability:
\begin{equation}
\langle \tau^m \rangle_{x_0} = m! \sum\limits_{n=0}^{\infty} u_n(x_0) \lambda_n^{-m} \int\limits_{-L}^L dx ~ u_n(x) \w(x) ;
\end{equation}

\item
from recurrence partial differential equations (PDEs)
\begin{equation}
\L^*_{x_0} \langle \tau^m \rangle_{x_0} = - m \langle \tau^{m-1} \rangle_{x_0} ,
\end{equation}
with Dirichlet boundary conditions \cite{Darling53}.  
\end{enumerate}
In what follows, we focus on the mean exit time $\langle
\tau\rangle_{x_0}$, the moment-generating function $\tilde{q}(x_0,s)$,
and the survival probability $S(x_0,t)$ for harmonically trapped
particles.

\subsection{Mean exit time}
\label{sec:mean}

The mean exit times of diffusive processes were studied particularly
well because of their practical importance and simpler mathematical
analysis (see \cite{Redner,Metzler,Weiss81,Szabo80,Weiss86} and
references therein).  In fact, the mean exit time,
\begin{equation}
\langle \tau \rangle_{x_0} = \int\limits_0^\infty dt ~ t~ q(x_0,t) = \int\limits_0^\infty dt~ S(x_0,t),
\end{equation}
satisfies the simpler equation than the time-dependent PDE
(\ref{eq:BFP}):
\begin{equation}
\L^*_{x_0} \langle \tau \rangle_{x_0} = -1 ,
\end{equation}
with Dirichlet boundary conditions at $x_0 = \pm L$.  The double
integration and imposed boundary conditions yield \cite{Hanggi90}%
\footnote{
In \cite{Hanggi90}, the sign minus in front of $U(z)$ in the
second integral in the numerator of the first term in Eq. (7.7) is
missing.}
\begin{equation}
\fl
\langle \tau \rangle_{x_0} = \frac{1}{D} \Biggl\{ \Biggl[\biggl(\int\limits_{-L}^L \frac{dx}{\w(x)}\biggr)^{-1} 
\int\limits_{-L}^L \frac{dx}{\w(x)} \int\limits_0^x dx' ~\w(x') \Biggr] \int\limits_{-L}^{x_0} \frac{dx}{\w(x)}  
 -  \int\limits_{-L}^{x_0} \frac{dx}{\w(x)} \int\limits_0^x dx'~ \w(x')\Biggr\} . 
\end{equation}
Substituting $w(x)$ from Eq. (\ref{eq:weight}), one gets
\begin{eqnarray}
\label{eq:tmean_1d_gen}
\langle \tau \rangle_{x_0} &=& \frac{L^2}{D} \frac{\sqrt{\pi}}{2\kappa} \Biggl\{ \frac{\erf(i\sqrt{\kappa}(x_0/L-\varphi)) + \erf(i\sqrt{\kappa}(1+\varphi))}
{\erf(i\sqrt{\kappa}(1-\varphi)) + \erf(i\sqrt{\kappa}(1+\varphi))} \\
\nonumber
&\times& \hspace*{-2mm} \int\limits_{\sqrt{\kappa}(-1-\varphi)}^{\sqrt{\kappa}(1-\varphi)} \hspace*{-2mm} dz ~ e^{z^2} \erf(z)
- \hspace*{-2mm} \int\limits_{\sqrt{\kappa}(-1-\varphi)}^{\sqrt{\kappa}(x_0/L-\varphi)} \hspace*{-2mm} dz ~ e^{z^2} \erf(z)  \Biggr\} ,
\end{eqnarray}
where $\erf(z)$ is the error function, and $\kappa$ and $\varphi$ are
two dimensionless parameters characterizing the trapping harmonic
potential and the pulling constant force, respectively
\begin{equation}
\label{eq:kappa_varphi}
\kappa \equiv \frac{k L^2}{2k_B T}, \qquad  \varphi \equiv \frac{\x}{L} = \frac{F_0}{k L} .
\end{equation}
Throughout the paper, we consider $\varphi \geq 0$ while all the
results for $\varphi < 0$ can be obtained by replacing $\varphi \to
-\varphi$ and $x_0 \to -x_0$.  For large $\kappa$ or $\varphi$, one
can use an equivalent representation (\ref{eq:tmean_1d_gen2}) provided
in \ref{sec:tmean_large_kappa}.

\begin{figure}
\begin{center}
\includegraphics[width=70mm]{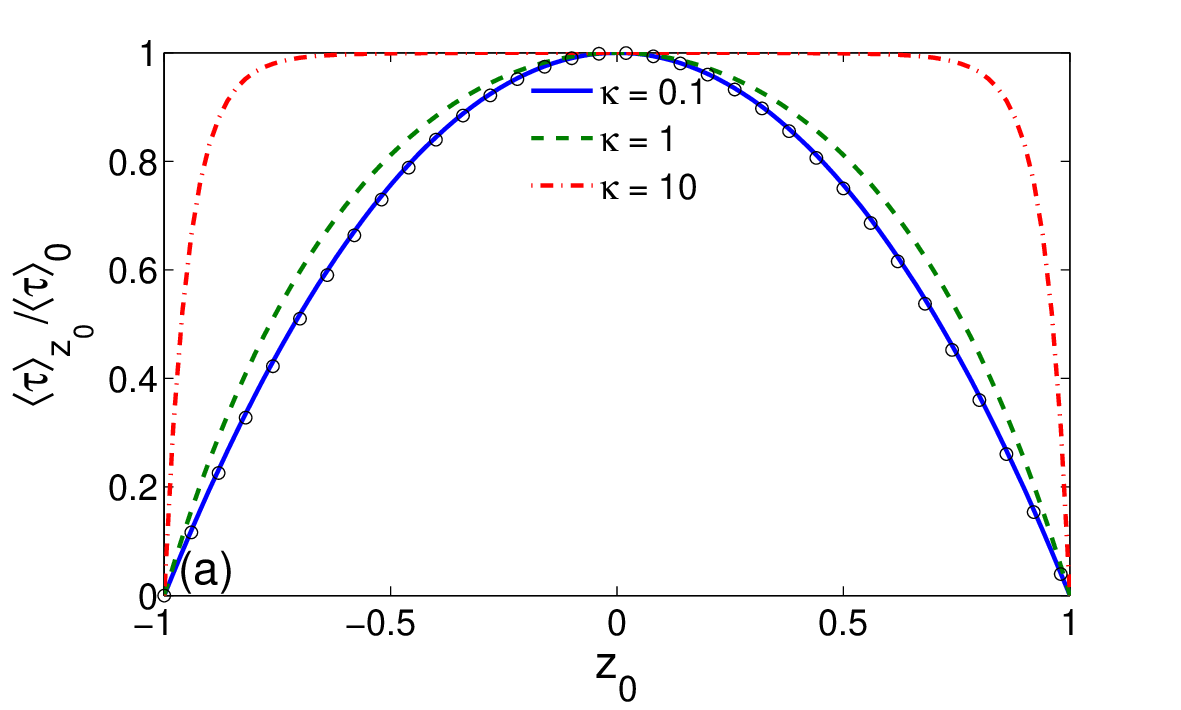}   % {tmean_1d_z0_kappa.eps}
\includegraphics[width=70mm]{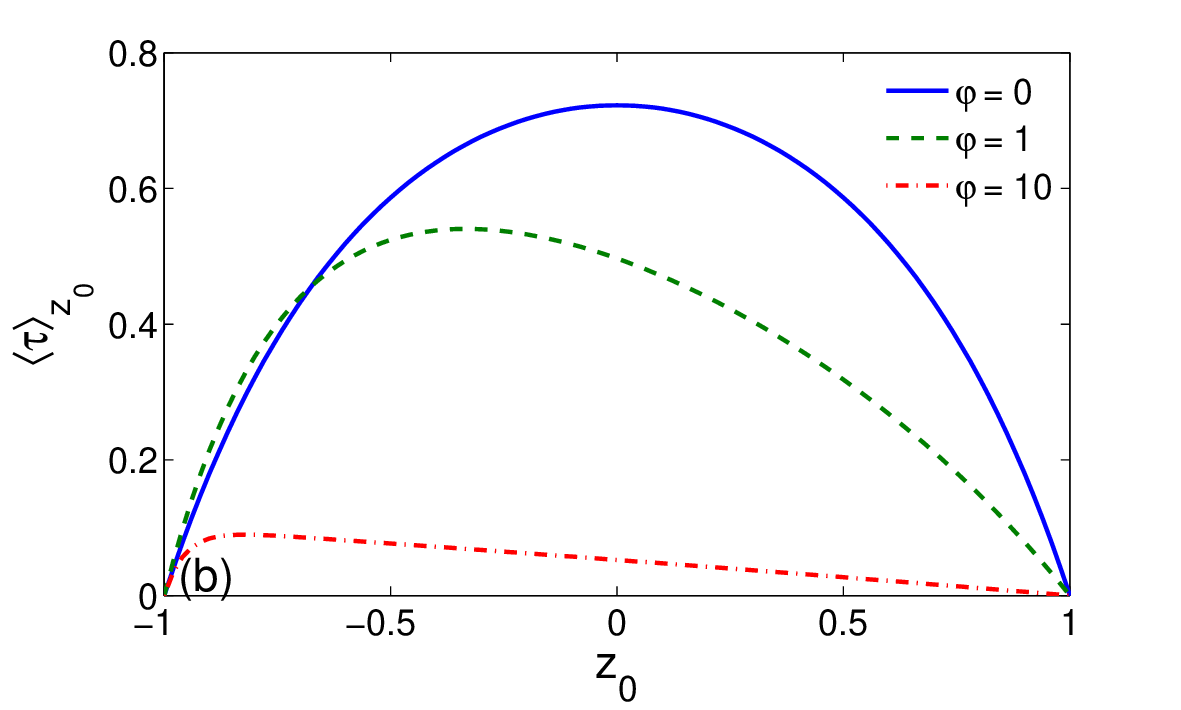}   % {tmean_1d_z0_varphi.eps}
\end{center}
\caption{
Mean exit time $\langle \tau\rangle_{z_0}$ as a function of $z_0 =
x_0/L$: for different $\kappa$ at fixed $\varphi = 0$ {\bf (a)} and
for different $\varphi$ at fixed $\kappa = 1$ {\bf (b)}.  The
timescale $L^2/D$ is set to $1$.  For plot {\bf (a)}, the mean exit
time is divided by its maximal value (at $z_0=0$) in order to rescale
the curves.  Circles indicate the mean exit time
$\frac{L^2}{2D}(1-z_0^2)$ without trapping ($k = 0$).}
\label{fig:tmean_1d_z0}
% plotted by A_Vert_article_fig_tmean_z0_kappa(),  A_Vert_article_fig_tmean_z0_varphi();
% calculated in tmean_1D.mw
% files with data: tmean_z0_kappa.txt (a), tmean_z0_varphi.txt (b)
\end{figure}

Several limiting cases are of interest:

$\bullet$ When $\varphi = 0$ (i.e., $F_0 = 0$),
Eq. (\ref{eq:tmean_1d_gen}) is reduced to
\begin{equation}
\langle \tau \rangle_{x_0} = \frac{L^2}{D} ~ \frac{\sqrt{\pi}}{2\kappa} 
\int\limits_{\sqrt{\kappa}~ x_0/L}^{\sqrt{\kappa}} dz~ e^{z^2}~ \erf(z) .
\end{equation}

$\bullet$ In the limit $k \to 0$, one gets a simpler expression
\begin{equation}
\langle \tau \rangle_{x_0} = \frac{L^2}{D \eta}~\biggl(1-x_0/L - 2\frac{e^{-\eta x_0/L} - e^{-\eta}}{e^{\eta} - e^{-\eta}} \biggr),   
\end{equation}
where $\eta = F_0 L/(k_B T)$ is another dimensionless parameter.  If
$F_0 = 0$, one retrieves the classical result for Brownian motion:
\begin{equation}
\label{eq:tmean_classic}
\langle \tau \rangle_{x_0} = \frac{L^2}{2D} \biggl(1 - (x_0/L)^2 \biggr) .
\end{equation}

$\bullet$ For small $\kappa$, the Taylor expansion of
Eq. (\ref{eq:tmean_1d_gen}) yields
\begin{equation}
\label{eq:tmean_1d_asympt}
\langle \tau \rangle_{x_0} \simeq \frac{L^2 - x_0^2}{2D} \biggl(1 + \kappa \frac{1 - 2\varphi (x_0/L) + (x_0/L)^2}{3} + O(\kappa^2) \biggr).
\end{equation}
We emphasize that the limits $\kappa\to 0$ and $k\to 0$ are not
equivalent because in the latter case, $\varphi\to\infty$ according to
Eq. (\ref{eq:kappa_varphi}).

$\bullet$ In the opposite limit of large $\kappa$, four cases can be
distinguished (see \ref{sec:tmean_large_kappa}):
\begin{equation}
\label{eq:tmean_1d}
\langle \tau \rangle_{x_0} \simeq \frac{L^2}{D} \cases{ \frac{\sqrt{\pi}~ e^{\kappa}}{4\kappa^{3/2}} & $(\varphi = 0)$, \\
\frac{\sqrt{\pi} ~ e^{\kappa(1-\varphi)^2}}{2\kappa^{3/2}(1-\varphi)} & $(0 < \varphi < 1)$, \\
\frac{1}{2\kappa} \ln \frac{\sqrt{\kappa}(1 - x_0/L)}{0.375\ldots}    & $(\varphi = 1)$, \\
\frac{1}{2\kappa} \ln \frac{\varphi - x_0/L}{\varphi-1}  & $(\varphi > 1)$, \\ }
\end{equation}
and the exponential growth in the first two relations is valid for any
$x_0$ not too close to $\pm L$.  Note that the limit of the second
asymptotic relation (for $0 < \varphi < 1$) as $\varphi \to 0$ is
different from the case $\varphi = 0$ by factor $2$.  In fact, when
$\varphi > 0$, it is much more probable to reach the right endpoint
than the left one, and $\langle \tau \rangle_{x_0}$ characterizes
mainly the exit through the right endpoint at large $\kappa$.  In
turn, when $\varphi = 0$, both endpoints are equivalent that doubles
the chances to exit and thus reduces by factor $2$ the mean exit time.
Note that the first two relations (up to a numerical prefactor)
can be obtained by the Kramers theory of escape from a potential well
\cite{Hanggi90,Kramers40}.  The last relation in
Eqs. (\ref{eq:tmean_1d}) can be retrieved from the last line of
Eq. (7.9) of Ref. \cite{Hanggi90}.

The behavior of the mean exit time $\langle \tau \rangle_{x_0}$ as a
function of the starting point $x_0$ is illustrated on
Fig. \ref{fig:tmean_1d_z0}.  The increase of $\kappa$ at fixed
$\varphi = 0$ transforms the spatial profile of the mean exit time
from the parabolic shape (\ref{eq:tmean_classic}) at $\kappa = 0$ to a
$\Pi$-shape at large $\kappa$ (Fig. \ref{fig:tmean_1d_z0}a).  In other
words, the dependence on the starting point becomes weak at large
$\kappa$.  At the same time, the height of the profile rapidly grows
with $\kappa$ according to Eq. (\ref{eq:tmean_1d}).  On the opposite,
the spatial profile becomes more skewed and sensitive to the starting
point as $\varphi$ increases at fixed $\kappa = 1$, while the height
is decreasing (Fig. \ref{fig:tmean_1d_z0}b).  As expected, the
constant force breaks the initial symmetry of the harmonic potential
and facilitates the escape from the trap.

\begin{figure}
\begin{center}
\includegraphics[width=70mm]{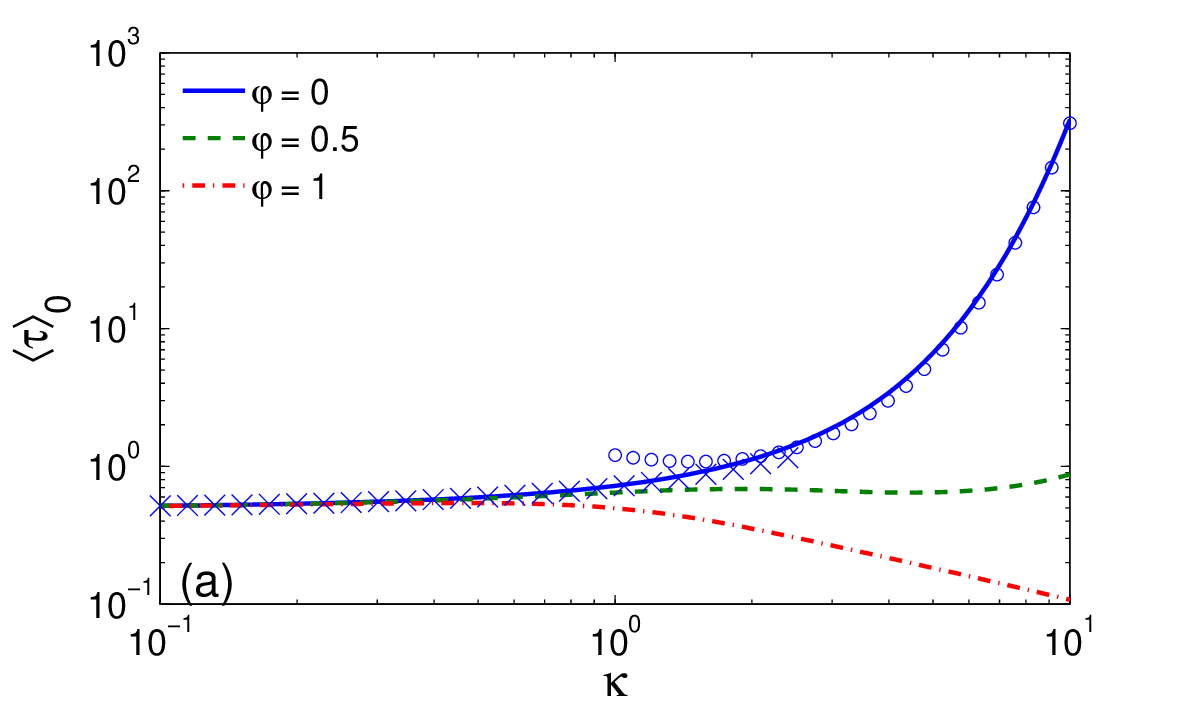}  % {tmean_1d_kappa.eps}
\includegraphics[width=70mm]{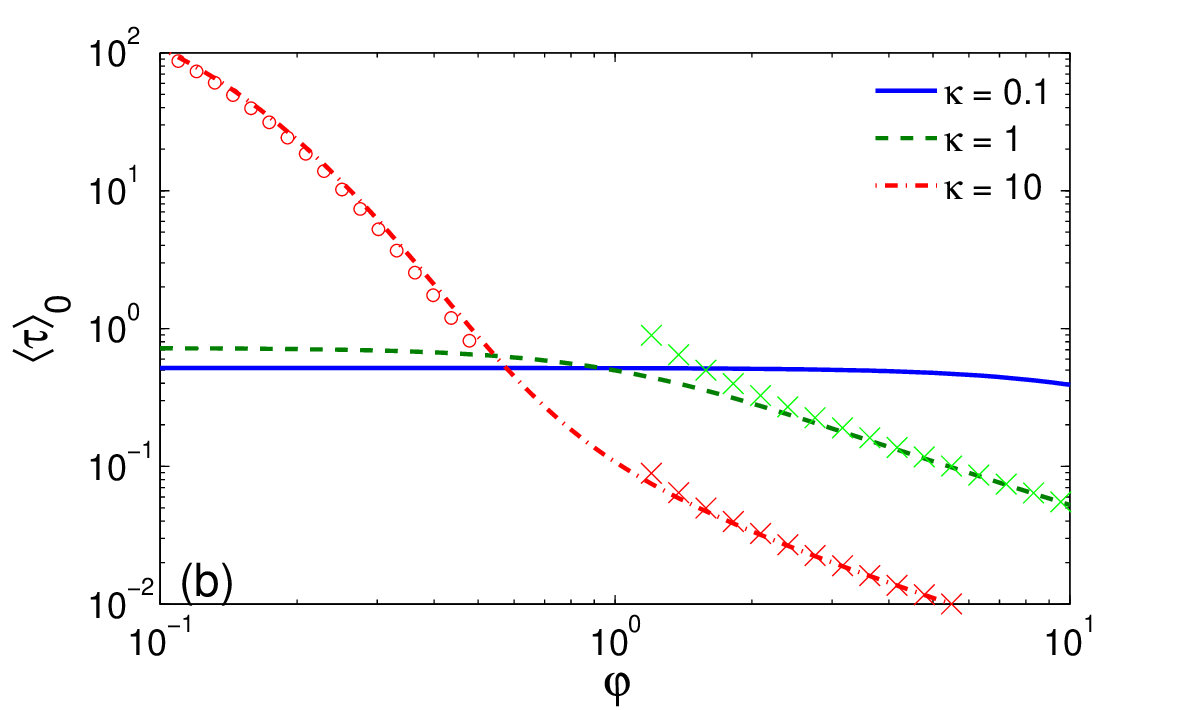}  % {tmean_1d_varphi.eps}
\end{center}
\caption{
Mean exit time $\langle \tau\rangle_0$ as a function of $\kappa$ for
fixed $\varphi$ {\bf (a)} and as a function of $\varphi$ for fixed
$\kappa$ {\bf (b)}.  The timescale $L^2/D$ is set to $1$.  On both
plots, circles show the exponential asymptotic relation in
Eq. (\ref{eq:tmean_1d}) for large $\kappa$ and $0 \leq \varphi < 1$.
On plot {\bf (a)}, crosses present the asymptotic
Eq. (\ref{eq:tmean_1d_asympt}) for small $\kappa$, to which the
next-order term, $2\kappa^2/45$, is added.  On plot {\bf (b)}, crosses
present the logarithmic asymptotic relation in Eq. (\ref{eq:tmean_1d})
for $\varphi > 1$.}
\label{fig:tmean_1d_kappa}
% plotted by A_Vert_article_fig_tmean_kappa(),  A_Vert_article_fig_tmean_varphi();
% calculated in tmean_1D.mw
% files with data: tmean_kappa.txt (a), tmean_varphi.txt (b)
\end{figure}

Figure \ref{fig:tmean_1d_kappa} shows how the mean exit time $\langle
\tau \rangle_0$ from the center varies with $\kappa$ and $\varphi$.
When there is no constant force ($\varphi = 0$), one observes a rapid
(exponential) growth at large $\kappa$, in agreement with
Eq. (\ref{eq:tmean_1d}) (shown by circles).  The presence of a
moderate constant force (with $0 < \varphi < 1$) slows down the
increase of the mean exit time.  For instance, at $\varphi = 0.5$,
$\langle \tau \rangle_0$ exhibits a broad minimum at intermediate
values of $\kappa$, but it resumes growing at larger values of
$\kappa$.  In turn, for $\varphi \geq 1$, there is no exponential
growth with $\kappa$, and the mean exit time slowly decreases, as
expected from Eq. (\ref{eq:tmean_1d}).  Since the constant force
shifts the minimum of the harmonic potential from $0$ to $\hat{x} =
F_0/k$, the border value $\varphi = 1$ corresponds to the minimum
$\hat{x}$ at the exit position ($\hat{x} = L$).  For $\varphi < 1$,
the harmonic potential keeps the particle away from the exit and thus
greatly increases the mean exit time.  In turn, for $\varphi > 1$, the
harmonic potential attracts the particle to $\hat{x}$ which is outside
the interval $[-L,L]$ and thus speeds up the escape.

Although we considered the FET from a symmetric interval $[-L,L]$ for
convenience, shifting the coordinate by $\x$ allows one to map the
original problem to the FET from a nonsymmetric interval $[-a,b]$ with
$a = L(1+\varphi)$ and $b = L(1-\varphi)$, with the starting point
$x_0$ being shifted by $L \varphi$ to vary from $-a$ to $b$.  As a
consequence, the choice of the symmetric interval $[-L,L]$ is not
restrictive, and all the results can be recast for a general interval
$[-a,b]$ by shifts.

\subsection{Survival probability}
\label{sec:spectral_1d}

The survival probability is fully determined by the eigenvalues and
eigenfunctions of the backward FP operator.  The eigenvalue equation
(\ref{eq:eigen}) reads
\begin{equation}
\label{eq:u_diffeq}
D u'' - (k/\gamma) (x - \x) u' + \lambda u = 0 .
\end{equation}
A general solution of this equation is well known \cite{Abramowitz}
\begin{equation}
\label{eq:u_KummerF}
\fl u(z) = c_1 M\left(-\frac{\alpha^2}{4\kappa}, \frac12, \kappa (z-\varphi)^2\right) 
 + c_2 (z-\varphi) M\left(-\frac{\alpha^2}{4\kappa}+\frac12, \frac32, \kappa (z-\varphi)^2\right) , 
\end{equation}
where $z = x/L$ is the dimensionless coordinate, $\lambda =
D\alpha^2/L^2$, $c_1$ and $c_2$ are arbitrary constants, and
\begin{equation}
\label{eq:KummerM}
M(a,b,z) = ~ _1F_1(a,b,z) = \sum\limits_{n=0}^\infty \frac{a^{(n)} z^n}{b^{(n)}  n!}  
\end{equation}
is the confluent hypergeometric function of the first kind (also known
as Kummer function), with $a^{(0)} = 1$ and $a^{(n)} = a(a+1)\ldots
(a+n-1) = \frac{\Gamma(a+n)}{\Gamma(a)}$, where $\Gamma(z)$ is the
gamma function.  The first and second terms in
Eq. (\ref{eq:u_KummerF}) are respectively symmetric and antisymmetric
functions with respect to $\varphi$.  

To shorten notations, we set
\begin{eqnarray}
\label{eq:mm}
m_{\alpha,\kappa}^{(1)}(z) &\equiv& M\left(-\frac{\alpha^2}{4\kappa}, \frac12, \kappa z^2\right) , \\
m_{\alpha,\kappa}^{(2)}(z) &\equiv& z M\left(-\frac{\alpha^2}{4\kappa}+\frac12, \frac32, \kappa z^2\right), 
\end{eqnarray}
so that
\begin{equation}
\label{eq:u_KummerF0}
u(z) = c_1 m_{\alpha,\kappa}^{(1)}(z-\varphi) + c_2 m_{\alpha,\kappa}^{(2)}(z-\varphi) .
\end{equation}

The Dirichlet boundary conditions read
\begin{eqnarray*}
c_1 m_{\alpha,\kappa}^{(1)}(-1-\varphi) + c_2 m_{\alpha,\kappa}^{(2)}(-1-\varphi) &=& 0  \qquad ({\rm at}~x_0 = -L),\\
c_1 m_{\alpha,\kappa}^{(1)}(1-\varphi) + c_2 m_{\alpha,\kappa}^{(2)}(1-\varphi) &=& 0   \qquad ({\rm at}~x_0 = L).
\end{eqnarray*}
In the special case $\varphi = 1$, one gets $c_1 = 0$, and the
eigenvalues are determined from the equation
$m_{\alpha,\kappa}^{(2)}(2) = 0$.  In general, for $\varphi \ne 1$,
one considers the determinant of the underlying $2\times 2$ matrix:
\begin{equation}
\label{eq:determ}
\D_{\alpha,\kappa,\varphi} = m_{\alpha,\kappa}^{(1)}(-1-\varphi) ~m_{\alpha,\kappa}^{(2)}(1-\varphi) 
 - m_{\alpha,\kappa}^{(2)}(-1-\varphi) ~m_{\alpha,\kappa}^{(1)}(1-\varphi) .
\end{equation}
Setting this determinant to $0$ yields the equation on $\alpha$:
\begin{equation}
\label{eq:eq_alpha}
\D_{\alpha_n, \kappa,\varphi} = 0 ,
\end{equation}
where $\alpha_n$ ($n = 0,1,2,\ldots$) denote all positive solutions of
this equation (for fixed $\kappa$ and $\varphi$).  The eigenfunctions
read then
\begin{equation}
\label{eq:u_KummerF2}
u_n(z) = \frac{\beta_n}{\sqrt{L}} \biggl[c_n^{(1)} m_{\alpha_n,\kappa}^{(1)}(z-\varphi) 
 - c_n^{(2)} m_{\alpha_n,\kappa}^{(2)}(z-\varphi) \biggr],
\end{equation}
where
\begin{equation}
\label{eq:cn}
c_n^{(1)} = m_{\alpha_n,\kappa}^{(2)}(1-\varphi),  \qquad c_n^{(2)} = m_{\alpha_n,\kappa}^{(1)}(1-\varphi), 
\end{equation}
and the normalization constant is 
\begin{equation}
\label{eq:beta_1d}
\beta_n^{-2} = \hspace*{-2mm} \int\limits_{-1-\varphi}^{1-\varphi} dz ~ e^{-\kappa z^2}
\biggl[c_n^{(1)} m_{\alpha_n,\kappa}^{(1)}(z) - c_n^{(2)} m_{\alpha_n,\kappa}^{(2)}(z) \biggr]^2 . 
\end{equation}

Multiplying Eq. (\ref{eq:u_diffeq}) by $\w(x)$ and integrating from
$a$ to $b$, one gets
\begin{equation}
\label{eq:us_int}
\int\limits_a^b dx ~ u(x)~ \w(x) = \frac{D}{\lambda} \bigl[u'(a) w(a) - u'(b) w(b)\bigr] .
\end{equation}
The derivative of the Kummer function can be expressed through Kummer
functions, in particular,
\begin{eqnarray}
\label{eq:mm_diff}
\partial_z m_{\alpha,\kappa}^{(1)}(z) &=& \frac{\alpha^2}{2\kappa z} \biggl(m_{\alpha,\kappa}^{(1)}(z) - m_{\sqrt{\alpha^2-4\kappa},\kappa}^{(1)}(z)\biggr) , \\
\partial_z m_{\alpha,\kappa}^{(2)}(z) &=& \left(2\kappa z^2 - 1 - \frac{\alpha^2}{2\kappa}\right) m_{\alpha,\kappa}^{(2)}(z)  
 + \left(2 + \frac{\alpha^2}{2\kappa}\right) m_{\sqrt{\alpha^2+4\kappa},\kappa}^{(2)}(z) , 
\end{eqnarray}
from which one gets explicit formulas for $u'_n(a)$ and $u'_n(b)$ and
thus for the integral in Eq. (\ref{eq:us_int}).  We get therefore
\begin{equation}
\label{eq:S_1d}
S(x_0,t) = \sum\limits_{n=0}^\infty w_n e^{-Dt \alpha_n^2/L^2} \biggl[c_n^{(1)} m_{\alpha_n,\kappa}^{(1)}(x_0/L-\varphi) 
- c_n^{(2)} m_{\alpha_n,\kappa}^{(2)}(x_0/L-\varphi) \biggr],  
\end{equation}
where
\begin{equation}
\label{eq:wn_1d_1}
w_n = \frac{\beta_n^2 e^{-\kappa}}{\alpha_n^2} \bigl[v(-1) - v(1) \bigr] ,
\end{equation}
with
\begin{equation}
v(z) = e^{2\kappa \varphi z} \biggl(c_n^{(1)}\partial_z m_{\alpha_n,\kappa}^{(1)}(z-\varphi) 
- c_n^{(2)} \partial_z m_{\alpha_n,\kappa}^{(2)}(z-\varphi)\biggr) .
\end{equation}
Taking the derivative with respect to time, one obtains the FET
probability density
\begin{equation}
\label{eq:qt_1d}
\fl q(x_0,t) = \frac{D}{L^2}\sum\limits_{n=0}^\infty w_n \alpha_n^2 e^{-Dt \alpha_n^2/L^2} \biggl[c_n^{(1)} m_{\alpha_n,\kappa}^{(1)}(x_0/L-\varphi) 
- c_n^{(2)} m_{\alpha_n,\kappa}^{(2)}(x_0/L-\varphi) \biggr]. 
\end{equation}

\begin{figure}
\begin{center}
\includegraphics[width=70mm]{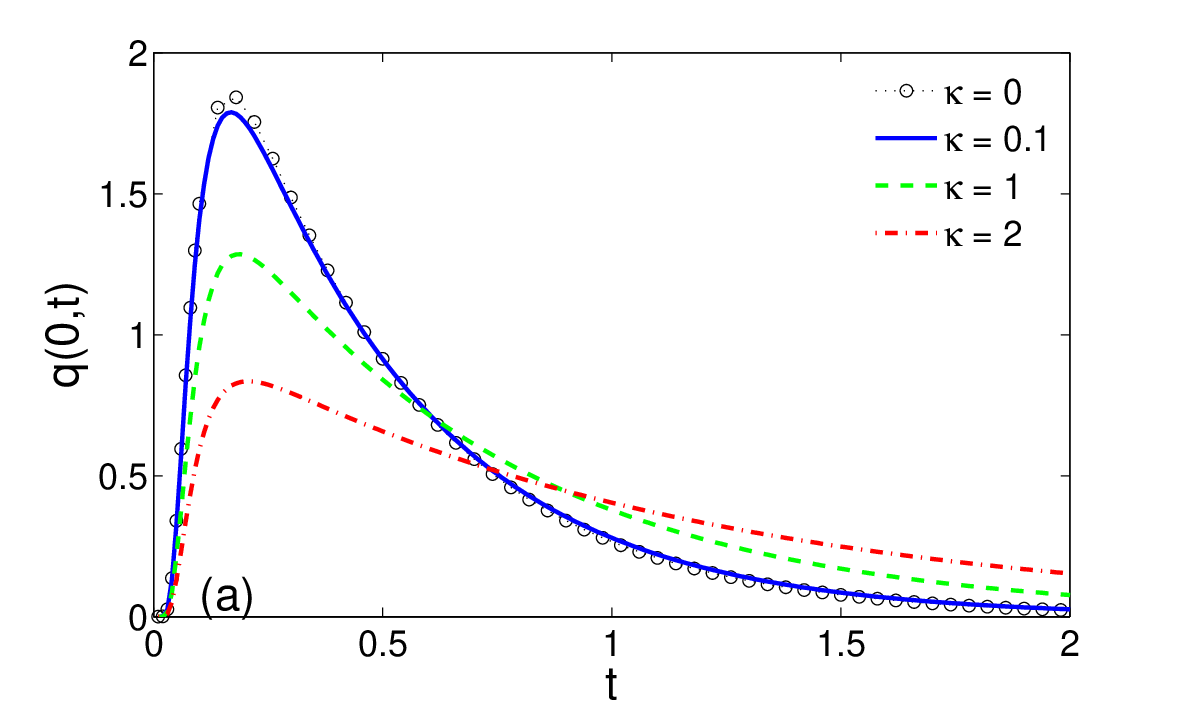}  % {qt_1d_kappa.eps}
\includegraphics[width=70mm]{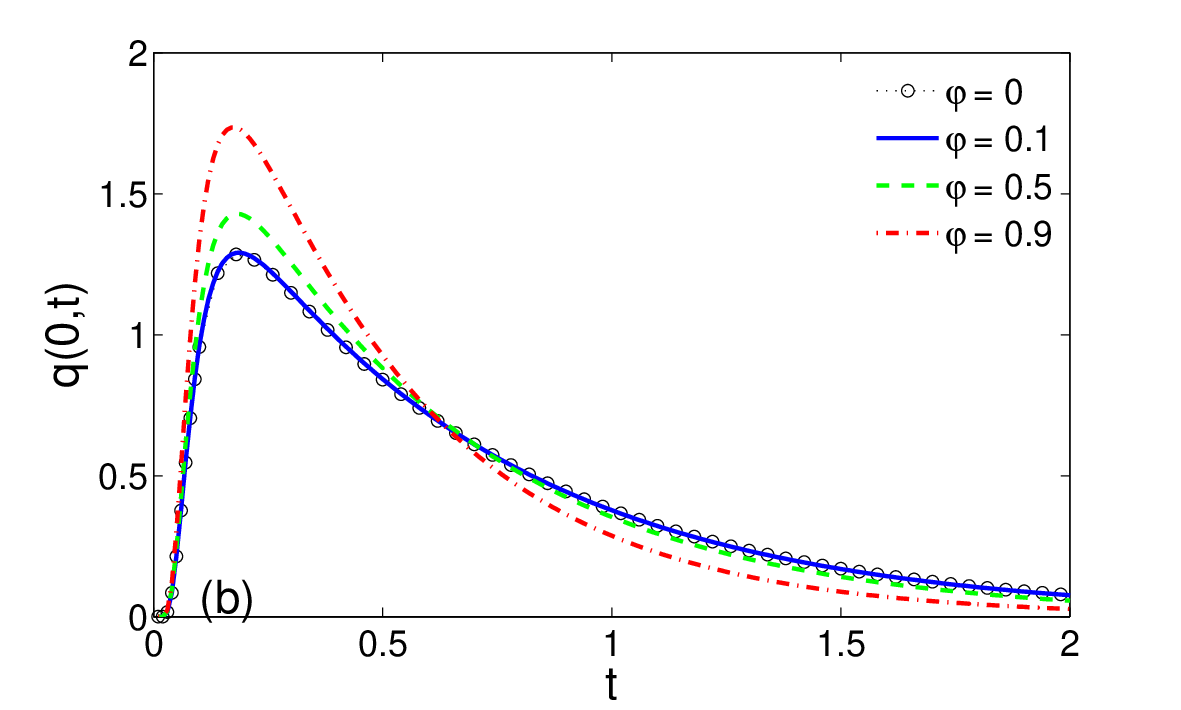}  % {qt_1d_varphi.eps}
\end{center}
\caption{
FET probability density $q(0,t)$ for several $\kappa$ at fixed
$\varphi = 0$ {\bf (a)} and for several $\varphi$ at fixed $\kappa =
1$ {\bf (b)}.  The timescale $L^2/D$ is set to $1$.  The spectral
decomposition (\ref{eq:qt_1d}) is truncated after $30$ terms. }
\label{fig:qt_1d_kappa}
% plotted by A_Vert_article_fig_qt_1d_kappa(), A_Vert_article_fig_qt_1d_varphi()
% calculated in survival_1D.mw
% files with data: spectral_1d_k0_phi0.txt , etc.
\end{figure}

In the limit $\kappa \to 0$, functions $m_{\alpha,\kappa}^{(1)}(z)$
and $m_{\alpha,\kappa}^{(2)}(z)$ approach $\cos(\alpha z)$ and
$\sin(\alpha z)$, respectively, so that eigenfunctions from
Eq. (\ref{eq:u_KummerF2}) become $u_n(z) = \frac{\beta_n}{\sqrt{L}}
\sin(\alpha(1-z))$, while the determinant in Eq. (\ref{eq:determ}) is
reduced to $\sin(2\alpha)$, from which $\alpha_n = \pi (n+1)/2$.  In
this limit, the dependence on $\varphi$ vanishes, and one retrieves
the classical result for Brownian motion
\begin{equation}
S(x_0,t) = 2 \sum\limits_{n=0}^\infty (-1)^n \frac{e^{-Dt \pi^2(n+1/2)^2/L^2}}{\pi(n+1/2)} ~ \cos(\pi(n+1/2) x_0/L)  .
\end{equation}
Only symmetric eigenfunctions with $\alpha_n = \pi (n+1/2)$ contribute
to this expression.

For centered harmonic potential ($\varphi = 0$),
Eq. (\ref{eq:eq_alpha}) is reduced to
\begin{equation}
m_{\alpha_n,\kappa}^{(1)}(1) ~m_{\alpha_n,\kappa}^{(2)}(1) = 0 ,
\end{equation}
which determines two sequences of zeros: $\alpha_{n,1}$ from
$m_{\alpha_{n,1},\kappa}^{(1)}(1) = 0$, and $\alpha_{n,2}$ from
$m_{\alpha_{n,2},\kappa}^{(2)}(1) = 0$.  As a consequence, one can
consider separately two sequences of symmetric and antisymmetric
eigenfunctions: $m_{\alpha_{n,1},\kappa}^{(1)}(z)$ and
$m_{\alpha_{n,2},\kappa}^{(2)}(z)$.  According to
Eq. (\ref{eq:S_spectral}), integration over arrival points removes all
the terms containing antisymmetric eigenfunctions.  This simpler
situation is considered as a particular case in Sec. \ref{sec:higher}.

Figure \ref{fig:qt_1d_kappa} illustrates the behavior of the
probability density $q(x_0,t)$.  For fixed $\varphi = 0$, an increase
of $\kappa$ increases the mean exit time and makes the distribution
wider.  Note that the most probable FET remains almost constant.  The
opposite trend appears for variable $\varphi$ at fixed $\kappa = 1$:
an increase of $\varphi$ diminishes the mean exit time and makes the
distribution narrower.  This is expected because a strong constant
force would drive the particle to one exit and dominate over
stochastic part.

\begin{figure}
\begin{center}
\includegraphics[width=70mm]{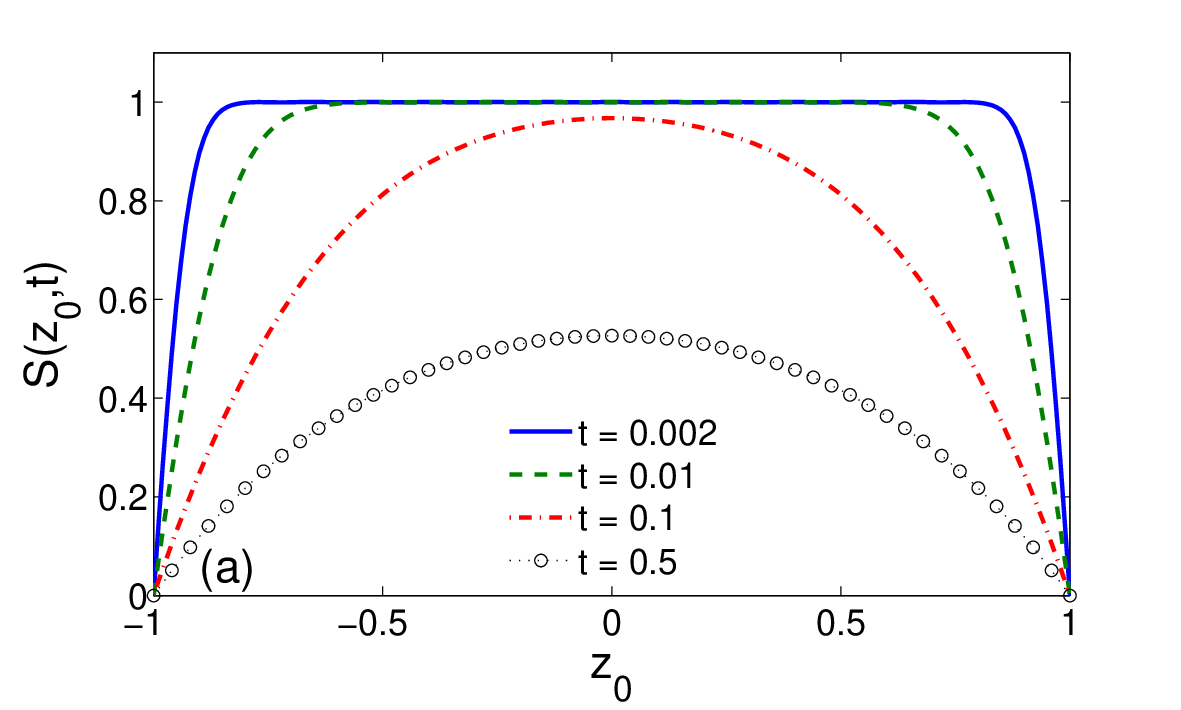}  % {St_1d_kappa.eps}
\includegraphics[width=70mm]{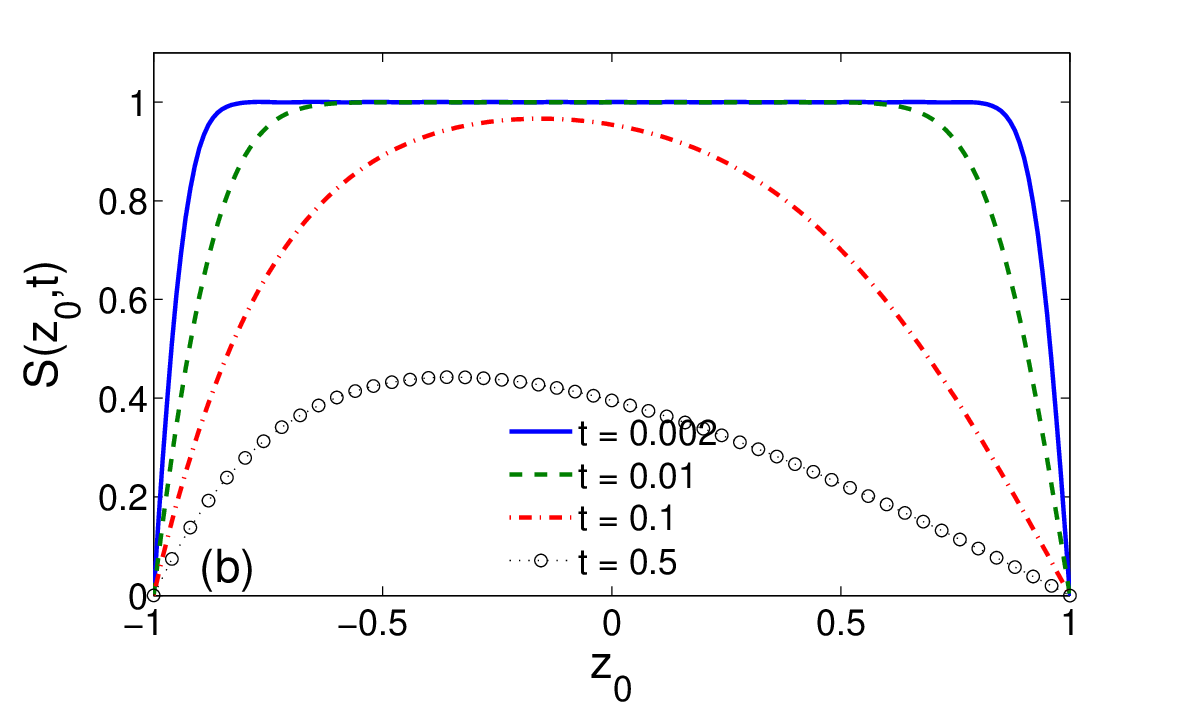}  % {St_1d_varphi.eps}
\end{center}
\caption{
Survival probability $S(x_0,t)$ as a function of the starting point
$z_0 = x_0/L$, with $\kappa = 1$, and $\varphi = 0$ {\bf (a)} and
$\varphi = 0.9$ {\bf (b)}.  The timescale $L^2/D$ is set to $1$.  The
spectral decomposition (\ref{eq:S_1d}) is truncated after $30$ terms.}
\label{fig:St_1d_kappa}
% plotted by A_Vert_article_fig_St_1d_kappa(), A_Vert_article_fig_St_1d_varphi()
% calculated in survival_1D.mw
% files with data: spectral_1d_k0_phi0.txt , etc.
\end{figure}

Figure \ref{fig:St_1d_kappa} shows the dependence of the survival
probability $S(x_0,t)$ on the starting point $x_0$.  At short times,
the survival probability is close to $1$ independently of $x_0$,
except for close vicinity of the endpoints.  As time increases,
$S(x_0,t)$ is progressively attenuated.  The spatial profile is
symmetric for centered harmonic potential ($\varphi = 0$), and skewed
to the left in the presence of a positive constant force ($\varphi =
0.9$): reaching the right endpoint is more probable due to the drift
by a constant force.

\subsection{Moment-generating function}
\label{sec:generating_1d}

Since any linear combination of functions in Eq. (\ref{eq:u_KummerF0})
satisfies Eq. (\ref{eq:FP_char}), with $s = -D\alpha^2/L^2$, one can
easily find the moment-generating function $\tilde{q}(x_0,s)$ by
imposing the boundary condition $\tilde{q}(\pm L,s) = 1$:
\begin{equation}
\label{eq:char_1d}
\tilde{q}(x_0,s) = \frac{A_{\alpha,\kappa,\varphi}^{(1)}}{\D_{\alpha,\kappa,\varphi}} m^{(1)}_{\alpha,\kappa}(x_0/L-\varphi) 
+ \frac{A_{\alpha,\kappa,\varphi}^{(2)}}{\D_{\alpha,\kappa,\varphi}} m^{(2)}_{\alpha,\kappa}(x_0/L-\varphi) ,
\end{equation}
where
\begin{eqnarray*}
A_{\alpha,\kappa,\varphi}^{(1)} &=& m^{(2)}_{\alpha,\kappa}(1-\varphi) - m^{(2)}_{\alpha,\kappa}(-1-\varphi) , \\
A_{\alpha,\kappa,\varphi}^{(2)} &=& m^{(1)}_{\alpha,\kappa}(-1-\varphi) - m^{(1)}_{\alpha,\kappa}(1-\varphi) , 
\end{eqnarray*}
and $\D_{\alpha,\kappa,\varphi}$ is defined by Eq. (\ref{eq:determ}).
Setting $a = L(1+\varphi)$ and $b = L(1-\varphi)$, one retrieves the
moment-generating function of the FET of an Ornstein-Uhlenbeck process
from an interval $[-a,b]$ reported in \cite{Borodin} (p. 548, 3.0.1),
in which Eq. (\ref{eq:char_1d}) is written more compactly in terms of
two-parametric family $S(\nu,a,b)$ of parabolic cylinder functions
(see \ref{sec:relations}).  For symmetric interval $[-a,a]$, a
similar expression for the moment-generating function was provided in
\cite{Darling53}.

It is worth noting that the probability density $q(x_0,t)$ could be
alternatively found by the inverse Laplace transform of
Eq. (\ref{eq:char_1d}).  For this purpose, one determines the poles
$s_n$ of $\tilde{q}(x_0,s)$ in the complex plane which are given by
zeros $\alpha_n$ of $\D_{\alpha,\kappa,\varphi}$ according to
Eq. (\ref{eq:eq_alpha}).  In other words, one has $s_n =
-D\alpha_n^2/L^2$, and the residue theorem yields
\begin{equation}
\fl q(x_0,t) = \frac{4\kappa D}{L^2} \sum\limits_{n=0}^\infty e^{-Dt\alpha_n^2/L^2} 
\biggl[\frac{A_{\alpha_n,\kappa,\varphi}^{(1)}}{\D'_{\alpha_n,\kappa,\varphi}} m^{(1)}_{\alpha_n,\kappa}(x_0/L-\varphi) 
 + \frac{A_{\alpha_n,\kappa,\varphi}^{(2)}}{\D'_{\alpha_n,\kappa,\varphi}} m^{(2)}_{\alpha_n,\kappa}(x_0/L-\varphi)\biggr], 
\end{equation}
where $\D'_{\alpha,\kappa,\varphi}$ denotes the derivative of
$\D_{\alpha,\kappa,\varphi}$ with respect to $s =
-\alpha^2/(4\kappa)$.  Comparing the above formula to
Eq. (\ref{eq:qt_1d}), one gets another representation for coefficients
$w_n$
\begin{equation}
\label{eq:wn_1d_2}
w_n = \frac{4\kappa}{\alpha_n^2} ~\frac{A_{\alpha_n,\kappa,\varphi}^{(1)}}{\D'_{\alpha_n,\kappa,\varphi}} ,
\end{equation}
where we used the identity $c_n^{(1)}
A_{\alpha_n,\kappa,\varphi}^{(2)} = - c_n^{(2)}
A_{\alpha_n,\kappa,\varphi}^{(1)}$, with $c_n^{(1,2)}$ from
Eq. (\ref{eq:cn}).  Two alternative representations (\ref{eq:wn_1d_1})
and (\ref{eq:wn_1d_2}) allow one to compute the normalization
coefficients $\beta_n$ without numerical integration in
Eq. (\ref{eq:beta_1d}).

\subsection{Higher-dimensional case}
\label{sec:higher}

In higher dimensions, we consider the FET of a multi-dimensional
Ornstein-Uhlenbeck process from a ball of radius $L$.  For centered
harmonic potential (i.e., $F_0 = 0$), the derivation follows the same
steps as earlier.  In fact, the integration of the probability density
$p(x,t|x_0,0)$ over the arrival point $x$ in the multi-dimensional
version of Eq. (\ref{eq:S_spectral}) removes the angular dependence of
the survival probability so that the eigenvalue equation is reduced to
the radial part
\begin{equation}
\label{eq:eigen_D}
\left(D\left[\partial^2_r + \frac{d-1}{r} \partial_r\right] - \frac{kr}{\gamma} \partial_r\right) u_n(r) + \lambda_n u_n(r) = 0 ,
\end{equation}
where $d$ is the space dimension.  In other words, we consider the FPT
of the radial Ornstein-Uhlenbeck process to the level $L$.  In turn,
the analysis for non-centered harmonic potential with $F_0 \ne 0$ is
much more involved in higher dimensions due to angular dependence, and
is beyond the scope of this review.

\subsubsection*{Survival probability.}

A solution of Eq. (\ref{eq:eigen_D}) is given by the Kummer function
which is regular at $r = 0$
\begin{equation}
\label{eq:un_D_int}
u_n(r) = \frac{\beta_n}{L^{d/2}} M\biggl(-\frac{\alpha_n^2}{4\kappa}, \frac{d}{2},\kappa (r/L)^2\biggr)   \qquad (n = 0,1,2,\ldots),
\end{equation}
where $\beta_n$ is the normalization factor:
\begin{equation}
\label{eq:beta_D_int}
\beta_n^{-2} = \int\limits_0^1 dz ~ z^{d-1} ~ e^{-\kappa z^2} \left[M\biggl(-\frac{\alpha_n^2}{4\kappa}, \frac{d}{2}, \kappa z^2\biggr)\right]^2 .
\end{equation}
The eigenvalues $\lambda_n = D\alpha_n^2/L^2$ are determined by the
positive zeros $\alpha_n$ of the equation
\begin{equation}
\label{eq:zeros_D_int}
M\biggl(-\frac{\alpha_n^2}{4\kappa}, \frac{d}{2}, \kappa\biggr) = 0.  
\end{equation}
Repeating the same steps as in Sec. \ref{sec:spectral_1d} yields the
spectral representation of the survival probability
\begin{equation}
S(r_0,t) = \sum\limits_{n=0}^\infty w_n ~ e^{-Dt\alpha_n^2/L^2}
M\left(-\frac{\alpha_n^2}{4\kappa}, \frac{d}{2}, \kappa (r_0/L)^2\right) ,
\end{equation}
where
\begin{equation} 
\label{eq:wn_D_int}
w_n = \frac{\beta_n^2 e^{-\kappa}}{2\kappa}~ M\left(-\frac{\alpha_n^2}{4\kappa}+1, \frac{d}{2}, \kappa\right),
\end{equation}
and we used the identity
\begin{equation} 
\int\limits_0^1 dz ~ z^{d-1} ~ e^{-\kappa z^2}~ M\left(-\frac{\alpha_n^2}{4\kappa}, \frac{d}{2}, \kappa z^2\right)  
 = \frac{e^{-\kappa}}{2\kappa} M\left(-\frac{\alpha_n^2}{4\kappa}+1 , \frac{d}{2}, \kappa\right). 
\end{equation}
The FET probability density is then
\begin{equation}
\label{eq:qt_D_int}
q(r_0,t) = \frac{D}{L^2} \sum\limits_{n=0}^\infty w_n ~ \alpha_n^2 ~ e^{-Dt\alpha_n^2/L^2}
M\left(-\frac{\alpha_n^2}{4\kappa}, \frac{d}{2}, \kappa (r_0/L)^2\right) .
\end{equation}

In the limit $\kappa \to 0$, one can use the identity (see \ref{sec:relations})
\begin{equation}
\fl
\lim\limits_{\kappa\to 0} M\left(-\frac{\alpha^2}{4\kappa},\frac{d}{2}, \kappa z^2\right) 
= \Gamma(d/2) \frac{J_{d/2 - 1}(\alpha z)}{(\alpha z/2)^{d/2-1}} = \cases{ \cos(\alpha z) & $(d = 1)$ \\
J_0(\alpha z) & $(d=2)$ \\  \frac{\sin(\alpha z)}{\alpha z} & $(d=3)$  \\ }
\end{equation}
to retrieve the classical results for Brownian motion (here $J_n(z)$
is the Bessel function of the first kind).  In particular, one
retrieves $\alpha_n = \pi(n+1/2)$ in one dimension and $\alpha_n =
\pi (n+1)$ in three dimensions (with $n = 0,1,2,\ldots$).  For the
one-dimensional case, we retrieved only the zeros of symmetric
eigenfunctions that contribute to the survival probability
(cf. discussion in Sec. \ref{sec:spectral_1d}).

\subsubsection*{Moment-generating function.}

The moment-generating function, obeying Eq. (\ref{eq:eigen_D}) with
$-s$ instead of $\lambda_n$, is
\begin{equation}
\label{eq:char_D_int}
\tilde{q}(r_0,s) = \frac{M\bigl(\frac{s L^2}{4\kappa D} , \frac{d}{2} , \kappa \frac{r_0^2}{L^2}\bigr)}
{M\bigl(\frac{s L^2}{4\kappa D} , \frac{d}{2} , \kappa\bigr)}  ,
\end{equation}
in agreement with \cite{Borodin} (p. 581, 2.0.1).  This function
satisfies the boundary condition $\tilde{q}(L,s) = 1$ and is regular
at $r_0 = 0$.  The Laplace inversion of this expression yields another
representation of the probability density
\begin{equation}
\label{eq:qt_D_int2}
q(r_0,t) = \frac{4\kappa D}{L^2} \sum\limits_{n=0}^\infty e^{-Dt\alpha_n^2/L^2} 
\frac{M\bigl(-\frac{\alpha_n^2}{4\kappa} , \frac{d}{2} , \kappa \frac{r_0^2}{L^2}\bigr)}
{M'\bigl(-\frac{\alpha_n^2}{4\kappa} , \frac{d}{2} , \kappa\bigr)} ,
\end{equation}
where $M'(a,b,z)$ denotes the derivative of $M(a,b,z)$ with respect to
$a$.  Comparing this relation to Eq. (\ref{eq:qt_D_int}), the
coefficients $w_n$ from Eq. (\ref{eq:wn_D_int}) can also be identified
as
\begin{equation}
\label{eq:wn_D_int_2}
w_n = \frac{4\kappa}{\alpha_n^2~M'\bigl(-\frac{\alpha_n^2}{4\kappa} ; \frac{d}{2} ; \kappa\bigr)} .
\end{equation}
As mentioned above, two expressions (\ref{eq:wn_D_int},
\ref{eq:wn_D_int_2}) for $w_n$ can be used to compute the
normalization constants $\beta_n$ without numerical integration in
Eq. (\ref{eq:beta_D_int}).

\subsubsection*{Mean exit time.}

Following the same steps as in Sec. \ref{sec:mean}, one gets the mean
exit time for the higher-dimensional case
\begin{equation}
\langle \tau \rangle_{r_0} = \frac{L^2}{D} ~ \frac{1}{\kappa}\int\limits_{\sqrt{\kappa} r_0/L}^{\sqrt{\kappa}} dr_1~ r_1^{1-d} e^{r_1^2} 
\int\limits_0^{r_1} dr_2 ~ r_2^{d-1} e^{-r_2^2} ,
\end{equation}
where we imposed Dirichlet boundary condition at $r_0 = L$ and the
regularity condition at $r_0 = 0$.  In the limit $\kappa \to 0$, one
retrieves the classical result $\langle \tau \rangle_{r_0} = (L^2 -
r_0^2)/(2dD)$.  In the opposite limit $\kappa \gg 1$, one gets
\begin{equation}
\label{eq:tmean_D_int}
\langle \tau \rangle_{r_0} \simeq \frac{L^2}{D} ~ \frac{\Gamma(d/2)~ e^{\kappa}}{4\kappa^{1+d/2}}  \qquad (\kappa \gg 1),
\end{equation}
which is applicable for any $r_0$ not too close to $L$.  The
behavior of the mean exit time for general spherically symmetric
potentials is discussed in \cite{Hanggi90}.

In \ref{sec:large_kappa}, the asymptotic behavior of the smallest
eigenvalue $\lambda_0 = D\alpha_0^2/L^2$ is obtained:
\begin{equation}
\label{eq:lambda1_D_int}
\lambda_0 \simeq \frac{D}{L^2} ~ \frac{4\kappa^{1+d/2}}{\Gamma(d/2)} ~ e^{-\kappa} \qquad (\kappa \gg 1),
\end{equation}
which is just the inverse of the above asymptotic relation for the
mean exit time.  While the first eigenvalue exponentially decays with
$\kappa$, the other eigenvalues linearly grow with $\kappa$ (see
\ref{sec:large_kappa}):
\begin{equation}
\lambda_n \simeq \frac{D}{L^2}~ 4\kappa n  \qquad (\kappa \gg 1)  .
\end{equation}
As a consequence, the gap between the lowest eigenvalue $\lambda_0$
and the next eigenvalue $\lambda_1$ grows linearly with $\kappa$.  For
$t \gg 1/\lambda_1$, the contribution of all excited eigenmodes
becomes negligible as compared to the lowest mode, and the first exit
time follows approximately an exponential law, $\P\{\tau > t\}
\simeq \exp(-t/\langle \tau\rangle)$, with the mean $\langle
\tau\rangle$ from Eq. (\ref{eq:tmean_D_int}).

\begin{figure}
\begin{center}
\includegraphics[width=70mm]{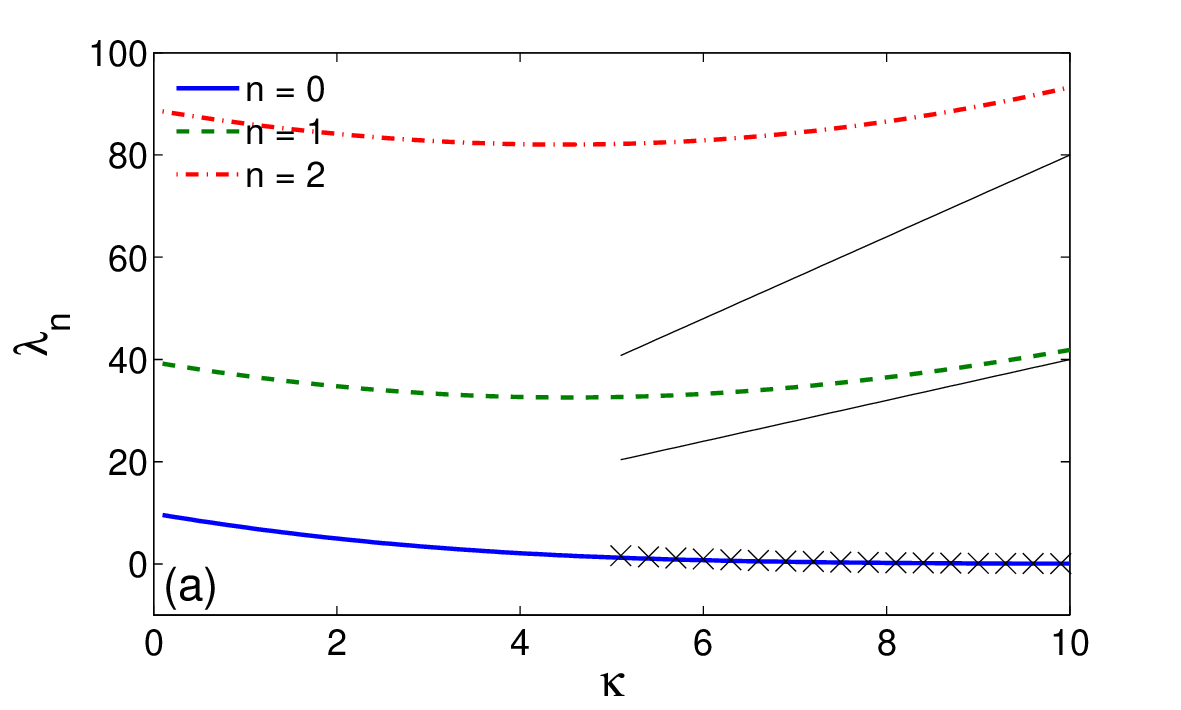}  % {lambda_kappa_3d_int.eps}
\includegraphics[width=70mm]{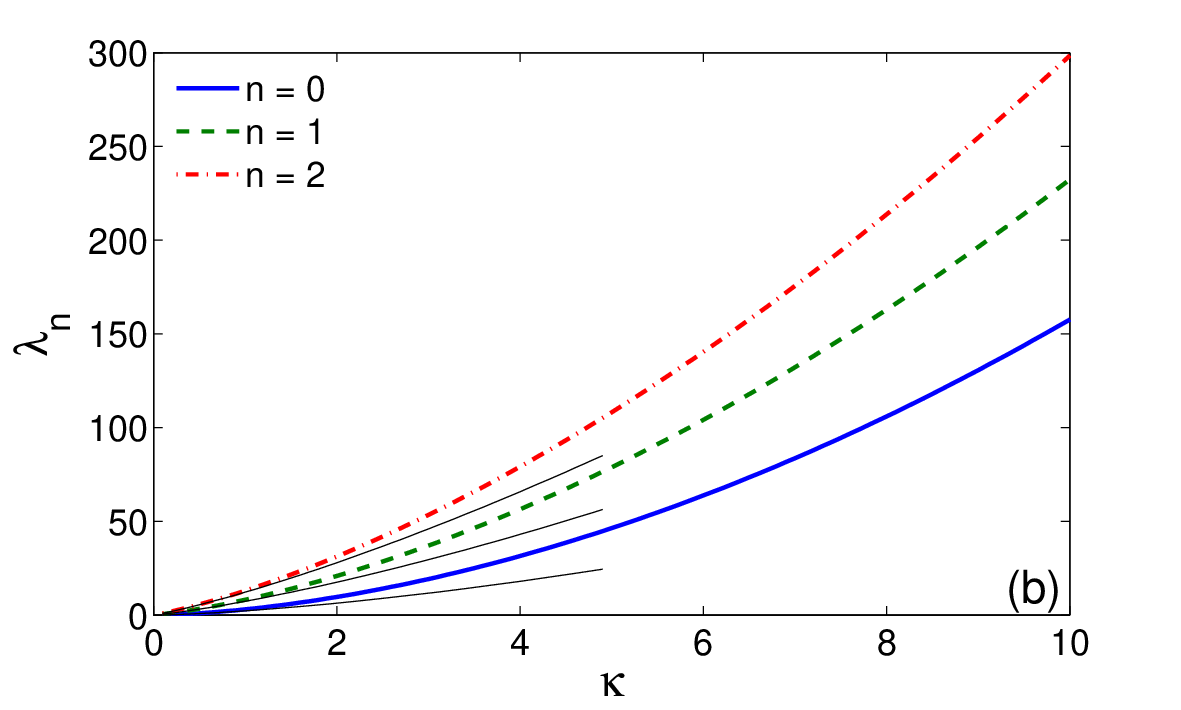}  % {lambda_kappa_3d_ext.eps}
\end{center}
\caption{
First three eigenvalues $\lambda_n$ of the Fokker-Planck operator as
functions of $\kappa$, for the interior problem {\bf (a)} and the
exterior problem {\bf (b)} in three dimensions.  The timescale $L^2/D$
is set to $1$.  On plot {\bf (a)}, crosses present the asymptotic
relation (\ref{eq:lambda1_D_int}) for the first eigenvalue $\lambda_0$
while thin solid lines indicate the asymptotic relation $4\kappa n$
for higher eigenvalues ($n = 1,2,\ldots$).  At $\kappa = 0$, one
retrieves the eigenvalues $\pi^2 (n+1)^2$ for Brownian motion.  On
plot {\bf (b)}, thin lines indicate the asymptotic behavior
(\ref{eq:alphan_D_ext}) at small $\kappa$. }
%$5.7832$, $30.4713$, and $74.8870$ which are squares of the first
%three roots of the Bessel function $J_0(z)$.}
\label{fig:alpha_1d_kappa}
%%%% plotted by A_Vert_article_fig_alpha_1d_kappa2(),  A_Vert_article_fig_alpha_2d_kappa()
% plotted by A_Vert_article_fig_alpha_int_kappa(),  A_Vert_article_fig_alpha_ext_kappa()
% calculated in survival_1D.mw
% files with data: spectral_1d_k0_phi0.txt , etc.
\end{figure}

We illustrate this behavior for three-dimensional case on
Fig. \ref{fig:alpha_1d_kappa}a which presents the first three
eigenvalues $\lambda_n$ as functions of $\kappa$.  Note that the
correction term to the asymptotic line for $\lambda_3$ is significant
even for $\kappa = 10$ (see \ref{sec:large_kappa} for details).  For
comparison, Fig. \ref{fig:alpha_1d_kappa}b shows the first three
eigenvalues for the exterior problem discussed in the next subsection.

\subsection{Exterior problem}
\label{sec:exterior}

For the exterior problem, when the process is started outside the
interval $[-L,L]$ (or outside the ball of radius $L$ in higher
dimensions), the FET is also referred to as the first passage time to
the boundary of this domain: $\tau = \inf\{ t>0 ~:~ |X(t)| < L\}$.
While the mean exit time and the probability distribution can be found
in a very similar way (see below), their properties are very different
from the earlier considered interior problem.  For the sake of
simplicity, we only consider the centered harmonic potential (i.e.,
$F_0 = 0$), although the noncentered case in one dimension can be
treated similarly.

In one dimension, the domain $(-\infty,-L)\cup (L,\infty)$ is split
into two disjoint subdomains so that $\tau$ is in fact the first
passage time to a {\it single} barrier, either at $x = L$ (if started
from $x_0 > L$), or at $x = -L$ (if started from $x_0 < -L$).  This
situation is described in \ref{sec:FPT}.

\subsubsection*{Mean exit time.}

Following the steps of Sec. \ref{sec:mean}, one obtains the mean exit
time
\begin{equation}
\langle \tau \rangle_{r_0} = \frac{L^2}{D} ~ \frac{1}{\kappa}\int\limits_{\sqrt{\kappa}}^{\sqrt{\kappa}~ r_0/L} dr_1~ r_1^{1-d} e^{r_1^2} 
\int\limits_{r_1}^{\infty} dr_2 ~ r_2^{d-1} e^{-r_2^2} ,
\end{equation}
where we imposed Dirichlet boundary condition at $r_0 = L$ and the
regularity condition at infinity.

For even dimensions $d$, the change of integration variables yields
the explicit formula
\begin{equation}
\langle \tau \rangle_{r_0} = \frac{L^2}{4D\kappa}\left[2\ln (r_0/L) + \sum\limits_{j=1}^{\frac{d}{2} -1} 
\frac{\Gamma(\frac{d}{2}\bigr)~ \kappa^{-j}}{j~ \Gamma(\frac{d}{2}-j\bigr)} (1 - (r_0/L)^{-2j}) \right]
\end{equation}
(we use the convention that $\sum\nolimits_{j=1}^n a_j$ is zero if $n
< 1$).  For instance, the mean exit time in two dimensions is
particularly simple:
\begin{equation}
\label{eq:tau_ext_2d}
\langle \tau \rangle_{r_0} = \frac{L^2}{2D\kappa} \ln (r_0/L)  \qquad (d = 2).
\end{equation}

For odd $d$, repeated integration by parts yields
\begin{eqnarray}
\fl 
\langle \tau \rangle_{r_0} &=& \frac{L^2}{4D} ~ \frac{1}{\kappa}\Biggl\{ 2\sqrt{\pi} \int\limits_{\sqrt{\kappa}}^{\sqrt{\kappa}~ z_0} 
dz ~ e^{z^2} \erfc(z) + \sum\limits_{j=1}^{\frac{d-3}{2}} \left(\frac{\Gamma\bigl(\frac{d}{2}\bigr)}{\Gamma\bigl(\frac{d}{2}-j\bigr)} - 
\frac{\Gamma\bigl(j+\frac{1}{2}\bigr)}{\Gamma\bigl(\frac{1}{2}\bigr)}\right) \frac{1 - z_0^{-2j}}{j~\kappa^j} \\
\nonumber
\fl
&+& e^{\kappa} \erfc(\sqrt{\kappa}) \sum\limits_{j=1}^{\frac{d-1}{2}} \Gamma\biggl(\frac{d}{2}-j\biggr) \kappa^{j-d/2}
- e^{\kappa z_0^2} \erfc(\sqrt{\kappa}~ z_0) 
\sum\limits_{j=1}^{\frac{d-1}{2}} \Gamma\biggl(\frac{d}{2}-j\biggr) (\kappa z_0^2)^{j-d/2} \Biggr\} ,
\end{eqnarray}
where $z_0 = r_0/L$, and $\erfc(z) = 1 - \erf(z)$.  Note that for $d =
1$, all terms vanish except the integral.

For large $r_0$ or large $\kappa$, the leading asymptotic term is
$\frac{L^2}{2D \kappa} \ln (r_0/L)$ for all dimensions (for odd
dimensions, this term comes from the integral).  When $r_0/L$
approaches $1$, the mean exit time vanishes as $c (r_0/L - 1)$, where
the prefactor $c$ depends on $\kappa$ and $d$.

When $\kappa\to 0$, the mean exit time diverges:
\begin{equation}
\langle \tau \rangle_{r_0} \simeq \frac{L^2}{D} ~ \frac{\Gamma\bigl(\frac{d}{2}\bigr)}{2(d-2)} \bigl(1 - (r_0/L)^{2-d}\bigr) \kappa^{-d/2}   \quad (d \ne 2) .
\end{equation}
(for $d = 2$, see Eq. (\ref{eq:tau_ext_2d})).  This divergence is
expected because, for the exterior problem, the mean exit time for
Brownian motion is infinite in all dimensions, irrespectively of its
recurrent or transient character.

Finally, the mean exit time for non-centered harmonic potential (i.e.,
$F_0 \ne 0$) in one dimension reads for $x_0 > L$ as
\begin{equation}
\langle \tau \rangle_{x_0} = \frac{L^2}{D} ~ \frac{\sqrt{\pi}}{2\kappa} \hspace*{-1mm} \int\limits_{\sqrt{\kappa}(1-\varphi)}^{\sqrt{\kappa}(x_0/L-\varphi)}
\hspace*{-2mm}  dz ~ e^{z^2} ~ \erfc(z) .
\end{equation}
In the limit of large $\kappa$, two asymptotic regimes are
distinguished:
\begin{enumerate}
\item
when $\varphi < 1$, the upper and lower limits of integration go to
infinity so that the mean exit time behaves as
\begin{equation}
\langle \tau \rangle_{x_0} \simeq \frac{L^2}{D} ~ \frac{1}{2\kappa} \ln \frac{x_0/L - \varphi}{1 - \varphi} ;
\end{equation}

\item
when $\varphi > 1$, the lower limit of integration goes to $-\infty$,
and the mean exit time exponentially diverges as
\begin{equation}
\langle \tau \rangle_{x_0} \simeq \frac{L^2}{D} ~ \frac{\sqrt{\pi}~ e^{\kappa(\varphi-1)^2}}{2\kappa^{3/2}(\varphi-1)}  .
\end{equation}
Both regimes are similar to that of the interior problem considered in
Sec. \ref{sec:mean}.

\end{enumerate}

\subsubsection*{Probability distribution.}

The moment-generating function $\tilde{q}(r_0,s)$ for the exterior
problem satisfies the same equation (\ref{eq:eigen_D}), with $-s$
instead of $\lambda_n$, as $\tilde{q}(r_0,s)$ from
Eq. (\ref{eq:char_D_int}) for the interior problem.  In order to
ensure the regularity condition at infinity (as $r_0\to\infty$), one
replaces $M(a,b,z)$ by the confluent hypergeometric function of
the second kind (also known as Tricomi function):
\begin{equation}
\label{eq:KummerU}
\fl U(a,b,z) = \frac{\Gamma(1-b)}{\Gamma(a-b+1)} M(a,b,z) + \frac{\Gamma(b-1)}{\Gamma(a)}~ z^{1-b} M(a-b+1,2-b,z)  
\end{equation}
(for integer $b$, this relation is undefined but can be extended by
continuity, see \ref{sec:computation}).  For $a>0$, the function
$U(a,b,z)$ vanishes as $z\to\infty$, in contrast to an exponential
growth of $M(a,b,z)$ according to Eqs. (\ref{eq:M_asympt},
\ref{eq:U_asympt}).  In turn, $U(a,b,z)$ exhibits non-analytic
behavior near $z = 0$, $U(a,b,z) \simeq
\frac{\Gamma(1-b)}{\Gamma(a-b+1)} + \frac{\Gamma(b-1)}{\Gamma(a)}
z^{1-b} + \ldots$, that limited its use for the interior problem.

The moment-generating function for the exterior problem is then
\begin{equation}
\label{eq:char_D_ext}
\tilde{q}(r_0,s) = \frac{U\bigl(\frac{s L^2}{4\kappa D} , \frac{d}{2} , \kappa \frac{r_0^2}{L^2}\bigr)}
{U\bigl(\frac{s L^2}{4\kappa D} , \frac{d}{2} , \kappa\bigr)}   \qquad (r_0 \geq L),
\end{equation}
in agreement with \cite{Borodin} (p. 581, 2.0.1).  

Denoting by $\alpha_n$ the positive zeros of the equation
\begin{equation}
U\left(-\frac{\alpha_n^2}{4\kappa} , \frac{d}{2} , \kappa\right) = 0 ,
\end{equation}
the inverse Laplace transform yields the FET probability density:
\begin{equation}
q(r_0,t) = \frac{4\kappa D}{L^2} \sum\limits_{n=0}^\infty e^{-D t\alpha_n^2/L^2} ~ 
\frac{U\bigl(- \frac{\alpha_n^2}{4\kappa} , \frac{d}{2} , \kappa \frac{r_0^2}{L^2}\bigr)}
{U'\bigl(-\frac{\alpha_n^2}{4\kappa} , \frac{d}{2} , \kappa\bigr)} ,
\end{equation}
where $U'(a,b,z)$ is the derivative with respect to $a$.  Its integral
over time is the survival probability:
\begin{equation}
\label{eq:S_D_ext1}
S(r_0,t) = \sum\limits_{n=0}^\infty w_n~ e^{-D t\alpha_n^2/L^2} ~ 
U\biggl(- \frac{\alpha_n^2}{4\kappa} , \frac{d}{2} , \kappa \frac{r_0^2}{L^2}\biggr)  ,
\end{equation}
with
\begin{equation}
w_n = \frac{4\kappa}{\alpha_n^2~ U'\bigl(-\frac{\alpha_n^2}{4\kappa} , \frac{d}{2} , \kappa\bigr)} .
\end{equation}

Alternatively, one can use the eigenvalues $\lambda_n =
D\alpha_n^2/L^2$ and the corresponding eigenfunctions
\begin{equation}
\label{eq:un_D_ext}
u_n(r) = \frac{\beta_n}{L^{d/2}}~ U\left(-\frac{\alpha_n^2}{4\kappa}, \frac{d}{2}, \kappa (r/L)^2\right), 
\end{equation}
where $\beta_n$ is the normalization factor:
\begin{equation}
\label{eq:beta_D_ext}
\beta_n^{-2} = \int\limits_1^\infty dz ~ z^{d-1} ~ e^{-\kappa z^2} \left[U\biggl(-\frac{\alpha_n^2}{4\kappa}, \frac{d}{2}, \kappa z^2\biggr)\right]^2 .
\end{equation}
The normalization factors $\beta_n$ diverge as $\kappa \to 0$.

Repeating the same steps as earlier, one retrieves the spectral
representation (\ref{eq:S_D_ext1}) of the survival probability with
\footnote{A misprint in Eq. (95) from the original paper is corrected.}
\begin{equation} \color{red}
%w_n = \frac{\beta_n^2 e^{-\kappa}}{2\kappa}~ U\left(-\frac{\alpha_n^2}{4\kappa}+1 , \frac{d}{2}, \kappa\right).
w_n = \frac12 \beta_n^2 e^{-\kappa}~ U\left(-\frac{\alpha_n^2}{4\kappa}+1 , \frac{d}{2} + 1, \kappa\right).
%= -\frac{\beta_n^2 e^{-\kappa}}{2\alpha_n^2}~ U\left(-\frac{\alpha_n^2}{4\kappa}, \frac{d}{2} + 1, \kappa\right).
\end{equation}
The asymptotic behavior of eigenvalues as $\kappa\to 0$ is discussed
in \ref{sec:large_kappa_ext}.

\subsection{Similarities and distinctions}

In spite of apparent similarities between the interior and the
exterior problems, there is a significant difference in spectral
properties of two problems.  This difference becomes particularly
clear in the limit $\kappa \to 0$ when the harmonic potential is
switched off (see Fig. \ref{fig:alpha_1d_kappa}).  For the interior
problem, the spectrum remains discrete and continuously approaches to
the spectrum of the radial Laplacian.  In this limit, one retrieves
the classical results for Brownian motion (e.g., $\alpha_n \to \pi
(n+1)/2$ as $\kappa\to 0$ in one dimension).  In turn, the Laplace
operator for the exterior problem has a continuum spectrum so that the
continuous transition from discrete to continuum spectrum as
$\kappa\to 0$ is prohibited.  In particular, all eigenvalues
$\lambda_n$ vanish as $\kappa\to 0$ (see \ref{sec:large_kappa_ext}).
In other words, the spectral properties for infinitely small $\kappa >
0$ and $\kappa = 0$ are drastically different.  One can see that the
asymptotic behavior of the eigenvalues $\lambda_n$ is quite different
for the interior and the exterior problems.

\section{Discussion}
\label{sec:discussion}

In this section, we discuss computational hints for confluent
hypergeometric functions (Sec. \ref{sec:hints}), three applications in
biophysics and finance (Sec. \ref{sec:tracking}, \ref{sec:bond},
\ref{sec:trading}), relation to the distribution of first crossing
times of a moving boundary by Brownian motion
(Sec. \ref{sec:crossing}), diffusion under quadratic double-well
potential (Sec. \ref{sec:double-well}), and further extensions
(Sec. \ref{sec:extensions}).

\subsection{Computational hints}
\label{sec:hints}

The probability distribution of first exit times involves confluent
hypergeometric functions $M(a,b,z)$ (for interior problem) and
$U(a,b,z)$ (for exterior problem).  For instance, the eigenvalues of
the Fokker-Planck operator are obtained through zeros $\alpha_n$ of
the equation $M(-\frac{\alpha_n^2}{4\kappa},\frac{d}{2},
\kappa) = 0$ or similar.  As a consequence, one needs to compute these
functions for large $|a| = \alpha_n^2/(4\kappa)$.  Although the series
in the definition (\ref{eq:KummerM}) of $M(a,b,z)$ converges for all
$z$, numerical summation becomes inaccurate for large $|a|$, and other
representations of confluent hypergeometric functions are needed.  In
\ref{sec:computation}, we discuss an efficient numerical scheme for
rapid and accurate computation of $M(a,b,z)$ for large $|a|$ and
moderate $z$ which relies on the expansion
(\ref{eq:KummerM_expansion}).  Moreover, we show that this scheme is
as well applicable for computing the derivative of $M(a,b,z)$ with
respect to $a$ which appears in Eq. (\ref{eq:qt_D_int2}) or similar
after the inverse Laplace transform.

For non-integer $b$, the Tricomi function $U(a,b,z)$ is expressed
through $M(a,b,z)$ by Eq. (\ref{eq:KummerU}) that allows one to apply
the same numerical scheme for the exterior problem in odd dimensions
$d$.  Although the Tricomi function for integer $b$ can be obtained by
continuation, the derivation of its rapidly converging representation
is more subtle.  In practice, one can compute $U(a,b,z)$ for an
integer $b$ by extrapolation of a sequence $U(a,b_\ve,z)$ computed for
non-integer $b_\ve$ approaching $b$ as $\ve \to 0$.  

In the case of large $z$ and moderate $|a|$, one can use
integral representations of $M(a,b,z)$ and $U(a,b,z)$ (see
\ref{sec:computation}).  The same algorithms can also be applied to
compute parabolic cylinder function $D_\nu(z)$ and Whittaker functions
(see \ref{sec:relations}).  The Matlab code for computing both Kummer
and Tricomi functions is available%
\footnote{
See \url{http://pmc.polytechnique.fr/pagesperso/dg/confluent/confluent.html}.}

\subsection{Single-particle tracking}
\label{sec:tracking}

The Langevin equation (\ref{eq:Langevin2}) can describe the thermal
motion of a small tracer in a viscous medium.  The Hookean force
$-kX(t)$ incorporates the harmonic potential of an optical tweezer
which are used to trap the tracer in a specific region of the medium
\cite{Rohrbach04}.  Optical trapping strongly diminishes the region
accessible to the tracer and thus enables to reduce the field of view
and to increase the acquisition rate up to few MHz
\cite{Lukic07,Franosch11,Huang11,Bertseva12,Grebenkov13b}.  At the same
time, trapping affects the intrinsic dynamics of the tracer and may
screen or fully remove its features at long times.  The choice of the
stiffness $k$ is therefore a compromise between risk of loosing the
tracer from the field of view (too small $k$) and risk of suppressing
important dynamical features (too large $k$).  The FET statistics can
then be used for estimating the appropriate stiffness due to a
quantitative characterization of escape events.  For instance, one can
choose the stiffness to ensure that the mean exit time strongly
exceeds the duration of experiment, or that the escape probability is
below a prescribed threshold.

Another interesting option consists in detecting events in which a
constant force is applied to the tracer.  In living cells, such events
can mimic the action of motor proteins that attach to the tracer and
pull it in one direction
\cite{Ashkin90,Kuo93,Brangwynne09,Sokolov12,Bressloff13}.  The
presence of a constant force facilitates the escape from the optical
trap while higher fraction of escape events (as compared to the case
without constant force) can be an indicator of such active transport
mechanisms.

Originally, the idea of fast escape in the case of comparable Hookean
and external forces was used to estimate the force generated by a
single protein motor \cite{Ashkin90}.  A ``trap and escape''
experiment consisted in trapping a single organelle moving along
microtubules at strong stiffness and then gradually reducing it until
the organelle escapes the trap.  Repeating such measurement, one can
estimate the ``escape power'' $kL$ as a measure of the driving force
$F_0$ when $\varphi = F_0/(kL) \sim 1$, where $L$ is the size of the
trap.  In this way, the driving force generated by a single
(presumably dynein-like) motor was estimated to be 2.6~pN
\cite{Ashkin90}.  This approximate but direct way of force measurement
relies on the drastic change in the mean exit time behavior at
$\varphi = 1$ according to Eq. (\ref{eq:tmean_1d}).

Interestingly, these mechanisms can even be detected from a single
trajectory.  When there is no constant force, the mean-square
displacement (MSD) of a trapped tracer, $\langle (\Delta X)^2\rangle$,
is known to approach the constant level $2k_B T/k$
\cite{Coffey,Desposito09,Grebenkov11}.  In other words, the square
root of the long-time asymptotic MSD determines the typical size
$\ell_k = \sqrt{2k_B T/k} = \sqrt{2D\tau_k}$ of the confining region
due to optical trapping, with $\tau_k = \gamma/k$.  Setting $L = q
\ell_k$, one gets $\kappa = q^2$, i.e., the dimensionless parameter
$\kappa$ can be interpreted as the squared ratio between the exit
distance $L$ and the characteristic size of the trap $\ell_k$.  The
above analysis showed that a tracer can rapidly reach levels which are
below or slightly above $\ell_k$.  However, significantly longer
explorations are extremely improbable.  In fact, according to
Eq. (\ref{eq:tmean_D_int}), the mean exit time for $\kappa \gg 1$ is
\begin{equation}
\label{eq:tau0}
\langle \tau \rangle_{0} \simeq \tau_k ~ \frac{\Gamma(d/2)~ e^{\kappa}}{2\kappa^{d/2}}  \qquad (\kappa \gg 1),
\end{equation}
where we set $L = q\ell_k = \sqrt{\kappa}~ \sqrt{2D\tau_k}$.  For
large enough $t$ (i.e., $t\gg \tau_k$), the contributions of all
excited eigenstates vanish, and the survival probability exhibits a
mono-exponential decay: $S(0,t)\simeq \exp(-t/\langle \tau
\rangle_{0})$, where we replaced the smallest eigenvalue $\lambda_0$
by $1/\langle \tau\rangle_0$ for $\kappa\gg 1$ according to
Eq. (\ref{eq:lambda1_D_int}), while $w_0 \approx 1$ as shown in
\ref{sec:large_kappa}.  In the intermediate regime $\tau_k
\ll t \ll \langle \tau \rangle_{0}$, the survival probability remains
therefore close to $1$.

A constant force $F_0$ pulling the tracer from the optical trap
strongly affects the mean exit time and the survival probability.  The
dimensionless parameter $\varphi$ from Eq. (\ref{eq:kappa_varphi}) is
the ratio between the new stationary position $\hat{x}$ of the
trajectory and the exit level $L =
\sqrt{\kappa}~ \ell_k$:
\begin{equation}
\varphi = \frac{F_0}{kL} = \frac{\hat{x}}{\ell_k \sqrt{\kappa}} = \frac{F_0 \sqrt{2D\tau_k}}{2k_B T \sqrt{\kappa}}  .
\end{equation}
For large $\varphi$, the mean exit time can be approximated according
to Eq. (\ref{eq:tmean_1d}) as $\langle \tau \rangle_0 \simeq
\tau_k \ln \frac{\varphi}{\varphi-1} \simeq \tau_k/\varphi$, i.e., it
becomes smaller than $\tau_k$, and much smaller than the mean exit
time from Eq. (\ref{eq:tau0}) without force.  As expected, exit events
would be observed much more often in the presence of strong constant
force.

\begin{figure}
\begin{center}
\includegraphics[width=70mm]{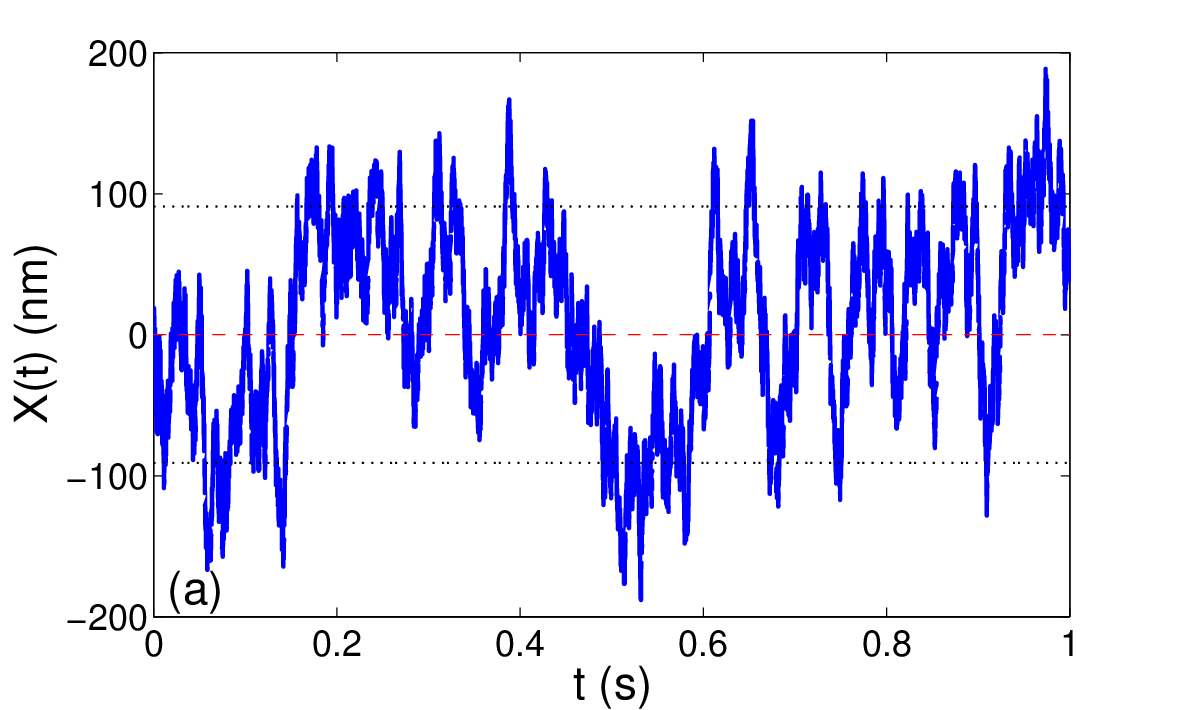}  % {traj1.eps}
\includegraphics[width=70mm]{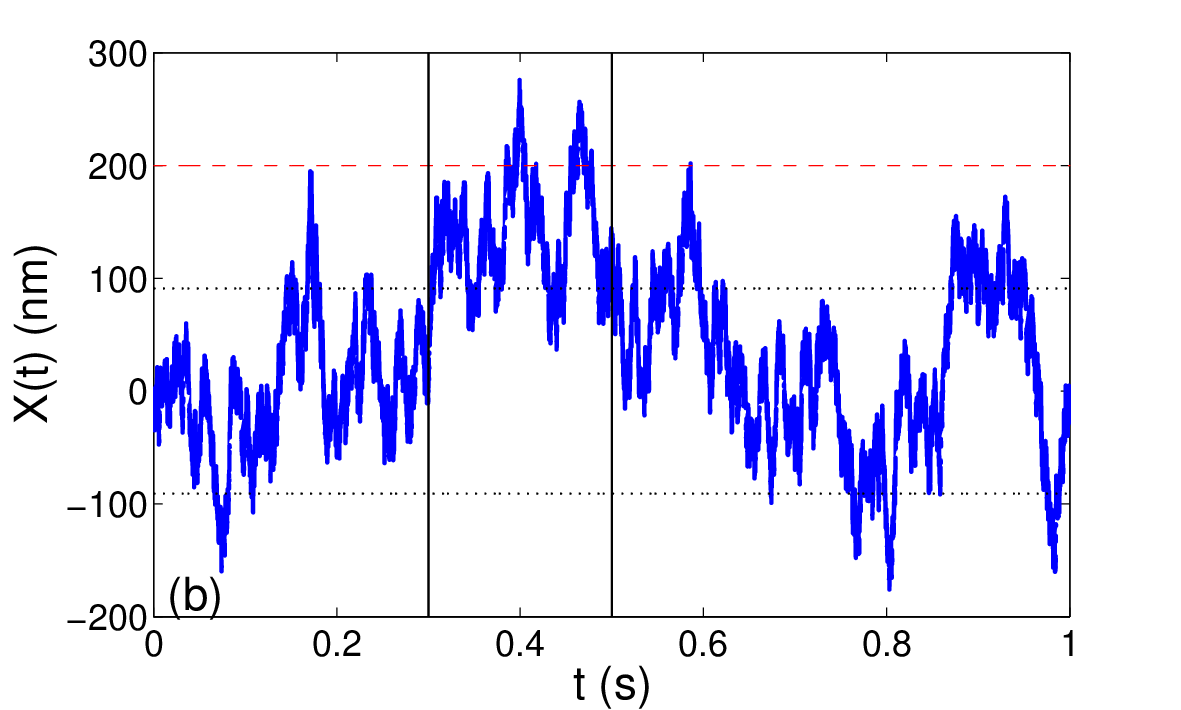}  % {traj2.eps}
\end{center}
\caption{
Two simulated trajectories of a spherical tracer submerged in water
under the optical trapping: {\bf (a)} no constant force; {\bf (b)}
constant force $F_0 = 0.2$~pN is applied between $0.3$ and $0.5$~s
(indicated by vertical lines).  Horizontal dotted lines indicate the
typical trapping size $\pm \ell_k = 91$~nm, while dashed red line
shows the stationary level $\hat{x}$ (equal to $0$ for zero force and
$200$~nm for $F_0 = 0.2$~pN).  The other parameters are provided in
the text.}
\label{fig:traj}
% A_Vert_simu_traj_fig();
\end{figure}

For a long acquired trajectory, one can characterize how often
different levels are reached.  Strong deviations from the expected
statistics (given by the survival probability) would suggest the
presence of a constant force.  To illustrate this idea, we simulate
the thermal motion of a spherical tracer of radius $a = 1~\mu$m
submerged in water and trapped by an optical tweezer with a typical
stiffness constant $k = 10^{-6}$~N/m \cite{Bertseva12,Grebenkov13b}.
The Stokes relation implies $\gamma = 6\pi a\eta_0 \approx 1.88\cdot
10^{-8}$~kg/s, from which the diffusion coefficient is $D = k_B
T/\gamma \approx 2.20\cdot 10^{-13}$~m$^2$/s at $T = 300$~K (with
$\eta_0 \approx 10^{-3}$~kg/m/s being the water viscosity).  The
characteristic trapping time is $\tau_k = \gamma/k
\approx 18.8$~ms, while the confinement length is $\ell_k = \sqrt{2k_B
T/k} \approx 91$~nm.  Figure \ref{fig:traj}a shows one simulated
trajectory of the tracer.  According to Eq. (\ref{eq:tmean_1d_gen}),
the mean exit times from the intervals $(-\ell_k,\ell_k)$ and
$(-2\ell_k,2\ell_k)$ are $27.2$~ms and $517$~ms, respectively.  For a
generated sample of duration 1~s, one observes multiple crossings of
levels $\pm \ell_k$ and only few crossings of levels $\pm 2\ell_k$.
For comparison, we generated another trajectory for which a constant
force $F_0 = 0.2$~pN (yielding $\varphi = 2.20/\sqrt{\kappa}$) is
applied between $0.3$~s and $0.5$~s (Fig. \ref{fig:traj}b).  Since
motor proteins exert forces which are typically tenfold higher
\cite{Ashkin90,Kuo93}, their effect is expected to be much stronger
and thus easier to detect.  The constant force reduces the mean exit
times to $9.8$~ms and $28$~ms, i.e., by factors $2.8$ and $18$,
respectively.  As expected, once the constant force is applied, the
tracer tends to reach the new stationary level $\x = 200$~nm so that
the trajectory crosses the level $\ell_k$ and remains above this level
for whole duration of the forced period.  Once the force is switched
off, the trajectory returns to its initial regime with zero mean.  One
can see that the use of FET statistics presents a promising
perspective for design and analysis of single-particle tracking
experiments, while Bayesian techniques can be further applied to get
more reliable results \cite{Bal04,Masson09}.  Note that the FPT
statistics have also been suggested as robust estimators of diffusion
characteristics \cite{Kenwright12} (see also
\cite{Condamin08}).  Another method relying on the time evolution of
the tracer probability distribution was proposed for simultaneously
extracting the restoring-force constant and diffusion coefficient
\cite{Lindner13}.

At the same time, we emphasize that this perspective needs further
analysis.  First, we focused on normal diffusion in a harmonic
potential while numerous single-particle tracking experiments
evidenced anomalous diffusion in living cells and polymer solutions
\cite{Bertseva12,Golding06,Szymanski09,Metzler09,Jeon11}.  Several
theoretical models have been developed to describe anomalous processes
such as continuous-time random walks (CTRW), fractional Brownian
motion (fBm), and generalized Langevin equation
\cite{Metzler00,Metzler04,Sokolov12,Bressloff13}.  While an extension
of the presented results is rather straightforward for CTRW
(Sec. \ref{sec:extensions}), the FPT problems for non-Markovian fBm or
generalized Langevin equation are challenging due to lack of
equivalent Fokker-Planck formulation.  Second, the quadratic profile
is an accurate approximation for optical trapping potential only for
moderate deviations from the center of the laser beam
\cite{Rohrbach04}, while the spatial profile can be more complicated
for strong deviations.  In other words, an accurate description of the
tracer escape may require more sophisticated analysis.  Finally, the
inference of constant forces from a single trajectory may present some
statistical challenges because different escape events can be
correlated.

\subsection{Adhesion bond dissociation under mechanical stress}
\label{sec:bond}

We briefly mention another biophysical example of bond dissociation.
Adhesion between cells or of cells to surfaces is mediated by weak
noncovalent interactions.  While a reversible bond between two
molecules can break spontaneously (due to thermal fluctuations), an
external force is needed to rupture multiple bonds that link two cells
together \cite{Bell78}.  The dynamics of bond rupture can be seen as
the first exit time problem in which exit or escape occurs when the
intermolecular distance exceeds an effective interaction radius.  Bell
suggested to apply the kinetic theory of the strength of solids to
describe the lifetime of a bond (i.e., the mean exit time) as
\begin{equation}
\label{eq:bond}
t_b = t_0 \exp[(E_b - r_b F_0)/(k_B T)], 
\end{equation}
where $E_b$ is the bond energy, $r_b$ is the range of the minimum of
the binding free energy, $F_0$ is the applied external force per bond,
and $t_0$ is the lifetime at the critical force $E_b/r_b$ at which the
minimum of the free energy vanishes \cite{Bell78}.  This relation
became a canonical description of adhesion bond dissociation under
force.
 
If the binding potential can be approximated as quadratic, then the
lifetime of a bond is precisely the mean exit time $\langle
\tau\rangle$ of a harmonically trapped particle.  In that case, the
second asymptotic relation in Eqs. (\ref{eq:tmean_1d}) implies the
quadratic dependence on the force, $\langle \tau \rangle \sim
e^{\kappa(1-\varphi)^2}$, where $\varphi = F_0/(k r_b)$, and $\kappa =
E_b/(k_B T) = kr_b^2/(2k_B T)$.  In other words, Eq. (\ref{eq:bond})
is retrieved only for weak forces when the quadratic term $\varphi^2$
can be neglected.  However, in the regime where the bond is most
likely to break, the applied force is large, and the mean exit time
may have completely different asymptotics (see, e.g., the last line of
Eqs. (\ref{eq:tmean_1d}) for $\varphi > 1$).  This discrepancy was
already outlined in Ref. \cite{Evans97}, in which the cases of a
harmonic potential and an inverse power law attraction were discussed,
and in Ref. \cite{Izrailev97} which presented molecular dynamics study
of unbinding and the related analysis of first exit times.  Other
effects such as the dependence of the bond strength and survival time
on the loading rate, were investigated both theoretically and
experimentally (see
\cite{Evans97,Izrailev97,Merkel99,Heymann00,Evans01,Butt05} and
references therein).

\subsection{Algorithmic trading}
\label{sec:trading}

Algorithmic trading is another field for applications of FETs.  In
algorithmic trading, a set of trading rules is developed in order to
anticipate the next price variation of an asset from its earlier
(historical) prices \cite{Box}.  Although the next price is random
(and thus unpredictable), one aims to catch some global or local
trends which can be induced by collective behavior of multiple traders
or macroeconomic tendences \cite{Covel,Clenow,Bouchaud}.  Many trading
strategies rely on the exponential moving average $\bar{p}_n$ of the
earlier prices $p_k$
\cite{Chan96,Moskowitz12,Asness13,Grebenkov14b}
\begin{equation}
\bar{p}_n = \lambda \sum\limits_{k=0}^{\infty} (1-\lambda)^k p_{n-k} ,
\end{equation}
where $0< \lambda \leq 1$ characterizes how fast the exponential
weights of more distant prices decay.  The difference between the
current price $p_n$ and the ``anticipated'' average price $\bar{p}_n$,
\begin{equation}
\delta_n \equiv p_n - \bar{p}_n = (1-\lambda)\sum\limits_{k=0}^{\infty} (1-\lambda)^{k} r_{n-k} ,  \qquad (r_n = p_n - p_{n-1}),
\end{equation}
can be seen as an indicator of a new trend.  For independent Gaussian
price variations $r_n$, writing $\delta_{n+1} = (1-\lambda) \delta_n +
(1-\lambda) r_{n+1}$, one retrieves Eq. (\ref{eq:AR}) for a discrete
version of an Ornstein-Uhlenbeck process, where $\hat{x} = \mu
(1-\lambda)/\lambda$ is related to the mean price variation $\mu$,
$\theta = -\ln(1-\lambda)$, and $\sigma = \sigma_0
\frac{(1-\lambda) \sqrt{2\theta}}{\sqrt{1-(1-\lambda)^2}}$ is
proportional to the standard deviation (volatility) $\sigma_0$ of
price variations.

The indicator $\delta_n$ can be used in both mean-reverting and trend
following strategies.  In the mean-reverting frame, if $\delta_n$
exceeds a prescribed threshold $L$, this is a trigger to sell the
asset at its actual (high) price, in anticipation of its return to the
expected (lower) level $\bar{p}_n$ in near future.  Similarly, the
event $\delta_n < -L$ triggers buying the asset.  In the opposite
trend following frame, the condition $\delta_n > L$ is interpreted as
the beginning of a strong trend and thus the signal to buy the asset
at its actual price, in anticipation of its further growth (similarly
for $\delta_n < -L$).  In other words, the same condition $\delta_n >
L$ (or $\delta_n < -L$) can be interpreted differently depending on
the empirical knowledge on the asset behavior.  Whatever the strategy
is used, the statistics of crossing of the prescribed levels $\pm L$
is precisely the FET problem.  Theoretical results in
Sec. \ref{sec:theory} can help to characterize durations between
buying and selling moments.  In particular, the choice of the
threshold $L$ is a compromise between execution of too frequent
buying/selling transactions (i.e., higher transaction costs) at small
$L$ and missing intermediate trends (i.e. smaller profits) at large
$L$.  We also note that Ornstein-Uhlenbeck processes often appear in
finance to model, e.g., interest rates (Vasicek model) and currency
exchange rates \cite{Hull,Vasicek77}.  A general frame of using
eigenfunctions for pricing options is discussed in \cite{Davydov03}.

\subsection{First crossing of a moving boundary by Brownian motion}
\label{sec:crossing}

The first exit time problem can be extended to time-evolving domains
\cite{Tuckwell84,Durbin85,Lerche86,Wang97,Kahale08}.  For instance,
one can investigate the first passage time of Brownian motion to a
time-dependent barrier $L(t)$, $\tau = \inf\{ t>0~:~ X(t) = L(t)\}$,
or the first exit time from a symmetric ``envelope'' $[-L(t),L(t)]$,
$\tau = \inf\{ t>0~:~ |X(t)| = L(t)\}$.  Although the survival
probability $S(x_0,t)$ satisfies the standard diffusion equation with
Dirichlet boundary condition, the boundary $L(t)$ evolves with time.
For a smooth $L(t)$, setting $S(x_0,t) = v(z,t)$ with a new space
variable $z = x_0/L(t)$ yields
\begin{equation}
\frac{\partial v(z,t)}{\partial t} = \frac{D}{L(t)^2}~ \partial^2_z v(z,t) - \frac{L'(t)}{L(t)}~ z~ \partial_z v(z,t),
\end{equation}
with Dirichlet boundary condition $v(\pm 1,t) = 0$ at two {\it fixed}
endpoints (here we focus on the exit time).  Setting a new time
variable $T = \ln (L(t)/L(0))$, the above equation can also be written
as the backward Fokker-Planck equation with time-dependent diffusion
coefficient $D(T) = D \frac{1}{L'(t) L(t)} = D
\frac{e^{-T}}{L(0)~ L'(L^{-1}(L(0)e^T))}$ and a centered harmonic potential:
\begin{equation}
\frac{\partial v(z,T)}{\partial T} = D(T) \partial^2_z v(z,T) -  z~ \partial_z v(z,T) .
\end{equation}
In higher dimensions, the second derivative $\partial^2_z$ is simply
replaced by the radial Laplace operator $\partial^2_r +
\frac{d-1}{r} \partial_r$.  

In general, the above equation does not admit explicit solutions.  A
notable exception is the case of square-root boundaries which has been
thoroughly investigated
\cite{Breiman66,Shepp67,Novikov71,Sato77,Salminen88,Novikov99}.
In fact, when $L(t) = \sqrt{2b(t + t_0)}$ (with $b > 0$ and $t_0 >
0$), one has $L'(t) L(t) = b$ so that $D(T)$ is independent of $T$ (or
$t$).  In other words, one retrieves the backward Fokker-Planck
problem (\ref{eq:FP_operator}, \ref{eq:FP}) with $\hat{x} = 0$,
$k/\gamma = 1$, $L = 1$, and $D$ replaced by $D/b$.  Its exact
solution is given by Eq. (\ref{eq:qt_D_int2}) for $d$-dimensional
case:
\begin{equation}
q(z_0,T) = 2\sum\limits_{n=0}^\infty e^{-2T\nu_n} \frac{M\bigl(-\nu_n, \frac{d}{2}, \frac{b z_0^2}{2D}\bigr)}{M'\bigl(-\nu_n, \frac{d}{2}, \frac{b}{2D}\bigr)} ,
\end{equation}
where $z_0 = r_0/L(0) = r_0/\sqrt{2bt_0}$ denotes the rescaled
starting point $r_0$, and $\nu_n = \alpha_n^2/(4\kappa)$ are zeros of
$M\bigl(-\nu,\frac{d}{2}, \frac{b}{2D}\bigr) = 0$.  Changing back $T$
to $t$, one gets
\begin{equation}
\label{eq:p_square}
p(z_0,t) = \frac{1}{t_0} \sum\limits_{n=0}^\infty (1 + t/t_0)^{-\nu_n - 1} ~ 
\frac{M\bigl(-\nu_n, \frac{d}{2}, \frac{b z_0^2}{2D} \bigr)}{M'\bigl(-\nu_n, \frac{d}{2}, \frac{b}{2D}\bigr)} ,
\end{equation}
This expression in a slightly different form was provided for $d = 1$
in \cite{Novikov99}.  Note also that the probability ${\mathbb P}\{
\sup_{1<t<T}(|W_t|/\sqrt{t}) < c\}$ admits a similar expansion
\cite{DeLong81}.  

In addition, Eq. (\ref{eq:char_D_int}) yields
\begin{equation}
t_0^{-\nu} \langle (\tau +t_0)^\nu \rangle = \langle e^{2\nu T} \rangle = 
\tilde{q}(z_0,-2\nu) = \frac{M\bigl(-\nu, \frac{d}{2}, \frac{b z_0^2}{2D} \bigr)}{M\bigl(-\nu, \frac{d}{2}, \frac{b}{2D}\bigr)}  ,
\end{equation}
from which one retrieves
\begin{equation}
\langle (\tau + t_0)^\nu \rangle = \frac{t_0^{\nu}}{M\bigl(-\nu, \frac{d}{2}, \frac{b}{2D}\bigr)}  \qquad ({\rm at}~z_0 = 0),
\end{equation}
that was reported for $d = 1$ in \cite{Shepp67}.  Note that the
$\nu$-th moment exists under the condition $\Re\{\nu\} < \nu_0$, as
clearly seen from Eq. (\ref{eq:p_square}).  In the special case $b =
D$, the square-root boundary $L(t) = \sqrt{2b(t+t_0)}$ grows in the
same way as the root-mean-square of Brownian motion $\sqrt{\langle
W_t^2\rangle} = \sqrt{2Dt}$.  Since $\nu_0 = 1$ at $b=D$, the mean
exit time is infinite.  More generally, the mean exit time is infinite
for broader envelopes ($b \geq D$) and finite for narrower envelopes
($b < D$), as expected.  The shift $t_0$ plays a minor role of a time
scale.

\subsection{Quadratic double-well potential}
\label{sec:double-well}

The above spectral approach can be extended to more complicated
trapping potentials.  As an example, we briefly describe diffusion
under double-well (or bistable) piecewise quadratic potential:
\begin{equation}
V(x) = \cases{ \frac12 k_1 (x + x_1)^2, & $x \leq 0$, \\
\frac12 k_2 (x - x_2)^2 + v_0,  & $x \geq 0$, } 
\end{equation}
where two minima are located at $-x_1$ and $x_2$ (with $x_1 > 0$ and
$x_2 > 0$), $k_1$ and $k_2$ are two spring constants, and $v_0 =
\frac12(k_1 x_1^2 - k_2 x_2^2)$ is a constant ensuring the continuity
of the potential at $x = 0$.  The resulting Langevin equation remains
linear, in contrast to other bistable potentials such as a quartic
potential (e.g., $V(x) = a x^4 + b x^2 + c x$).  The diffusive
dynamics under double-well potentials was thoroughly investigated by
using general theoretical tools (e.g. Kramers' theory
\cite{Hanggi90,Kramers40} or WKB approximation
\cite{Merzbacher,Griffiths,Caroli79}) and exactly solvable models (see
\cite{vanKampen77,Morsch79,Hongler82,Voigtlaender85,Jung85,Ivlev88,Kalmykov06}
and references therein).

For each semi-axis, an eigenfunction satisfies Eq. (\ref{eq:u_diffeq})
with the proper $k_i$.  However, neither Kummer, nor Tricomi function
is appropriate to represent the solution in this case.  In fact, the
Kummer function $M(a,1/2,z^2)$ rapidly grows at infinity, while the
Tricomi function $U(a,1/2,z^2)$ behaves as 
$\frac{\sqrt{\pi}}{\Gamma(a+1/2)} -
\frac{2\sqrt{\pi}}{\Gamma(a)} |z| + \ldots$ for small $z$, i.e., its
derivative is discontinuous at $0$.  A convenient representation can
still be obtained as a linear combination of two Kummer functions, in
which the rapid growth of these functions is compensated.  This is
precisely the case of parabolic cylinder functions $D_\nu(z)$ and
$D_\nu(-z)$ which vanish as $z\to\infty$ (resp., $z\to -\infty$) but
rapidly grow as $z\to -\infty$ (resp. $z\to\infty$) unless $\nu$ is a
nonnegative integer [see Eqs. (\ref{eq:CylinderD1}),
(\ref{eq:Dnu_large1}), (\ref{eq:Dnu_large2})].  An eigenfunction can
therefore be written as
\begin{equation}
u(x) = \cases{ c_1 e^{\kappa_1(x/x_1+1)^2/2} D_{\nu_1}\biggl(- \sqrt{2\kappa_1}(x/x_1+1) \biggr) , & $x \leq 0$, \\
c_2 e^{\kappa_2(x/x_2-1)^2/2} D_{\nu_2}\biggl(\sqrt{2\kappa_2}(x/x_2-1) \biggr),  & $x \geq 0$,  }
\end{equation}
where $\kappa_i = k_i x_i^2/(2k_B T)$, $\nu_i = \lambda
x_i^2/(2\kappa_i D)$, and $\lambda$, $c_1$, $c_2$ are determined by
normalization and two interface conditions at $x = 0$.  The continuity
of the eigenfunction at $x = 0$ can be satisfied by choosing
\begin{eqnarray*}
c_1 &=& \beta~ e^{\kappa_2/2} D_{\nu_2}(-\sqrt{2\kappa_2}), \qquad  c_2 = \beta~ e^{\kappa_1/2} D_{\nu_1}(-\sqrt{2\kappa_1}),
\end{eqnarray*}
where $\beta$ is a normalization constant.

The second interface condition is deduced from the orthogonality of
eigenfunctions with two weights $w_{1,2}$ from Eq. (\ref{eq:weight})
for positive and negative semi-axes,
\begin{equation*}
w_i(x) = \exp\bigl(\kappa_i[ 1 - (x/x_i \pm 1)^2]\bigr),
\end{equation*}
where plus (resp., minus) corresponds to $i=1$ (resp., $i=2$).  The
orthogonality imposes the interface condition
\begin{equation}
D w_1(0) u'(0^-) - D w_2(0) u'(0^+) = 0 ,
\end{equation}
where the same diffusion coefficient $D$ is assumed on both sides.
Since $w_i(0) = 1$, one retrieves the standard flux continuity
equation, $u'(0^-) = u'(0^+)$, yielding an equation determining the
eigenvalues $\lambda$:
\begin{eqnarray*}
\fl
&& x_1 D_{\nu_1}(-\sqrt{2\kappa_1}) \biggl[2\kappa_2 D_{\nu_2}(-\sqrt{2\kappa_2}) + \sqrt{2\kappa_2} D_{\nu_2+1}(-\sqrt{2\kappa_2})\biggr] + \\
&&  x_2 D_{\nu_2}(-\sqrt{2\kappa_2}) \biggl[2\kappa_1 D_{\nu_1}(-\sqrt{2\kappa_1}) + \sqrt{2\kappa_1} D_{\nu_1+1}(-\sqrt{2\kappa_1})\biggr] = 0 ,
\end{eqnarray*}
where we used the identity $\frac{\partial}{\partial z} D_\nu(z) =
\frac{z}{2} D_\nu(z) - D_{\nu+1}(z)$, and $\lambda$ appears in
$\nu_i = \lambda x_i^2/(2\kappa_i D)$.  The smallest eigenvalue
$\lambda = 0$ corresponds to the steady state.

The normalization constant $\beta$ is found according to
\begin{equation}
\label{eq:beta_DW}
\fl
\beta^{-2} = e^{\kappa_1+\kappa_2} \left[ \frac{x_1 D_{\nu_2}^2(-\sqrt{2\kappa_2})}{\sqrt{2\kappa_1}}  
\int\limits_{-\sqrt{2\kappa_1}}^{\infty} dz ~ D_{\nu_1}^2(z)
+ \frac{x_2 D_{\nu_1}^2(-\sqrt{2\kappa_1})}{\sqrt{2\kappa_2}} \hspace*{-1mm} \int\limits_{-\sqrt{2\kappa_2}}^\infty dz ~ D_{\nu_2}^2(z) \right] ,
\end{equation}
in which both integrals can be partly computed by using the identity
\cite{Gradshteyn}
\begin{equation}
\int\limits_0^\infty dz~ D_\nu^2(z) = \frac{\sqrt{\pi}}{2^{3/2}} ~\frac{\psi(\frac{1-\nu}{2}) - \psi(-\frac{\nu}{2})}{\Gamma(-\nu)} ,
\end{equation}
where $\psi(z) = \Gamma'(z)/\Gamma(z)$ is the digamma function.  The
lowest eigenfunction corresponding to $\lambda_0 =0$, is constant,
$u_0(x) = \beta_0$, with
\begin{equation*}
\beta_0^{-2} = \frac{\sqrt{\pi}}{2} \left[\frac{x_1 e^{\kappa_1} (1 + \erf(\sqrt{\kappa_1}))}{\sqrt{\kappa_1}} + 
\frac{x_2 e^{\kappa_2} (1 + \erf(\sqrt{\kappa_2}))}{\sqrt{\kappa_2}}\right].
\end{equation*}
As a consequence, one retrieves the equilibrium Boltzmann-Gibbs
distribution, $p_{\rm eq}(x) = p(x,\infty|x_0,0) = u_0(x_0) u_0(x)
\w(x) = \beta_0^2 \w(x)$.

\begin{figure}
\begin{center}
\includegraphics[width=70mm]{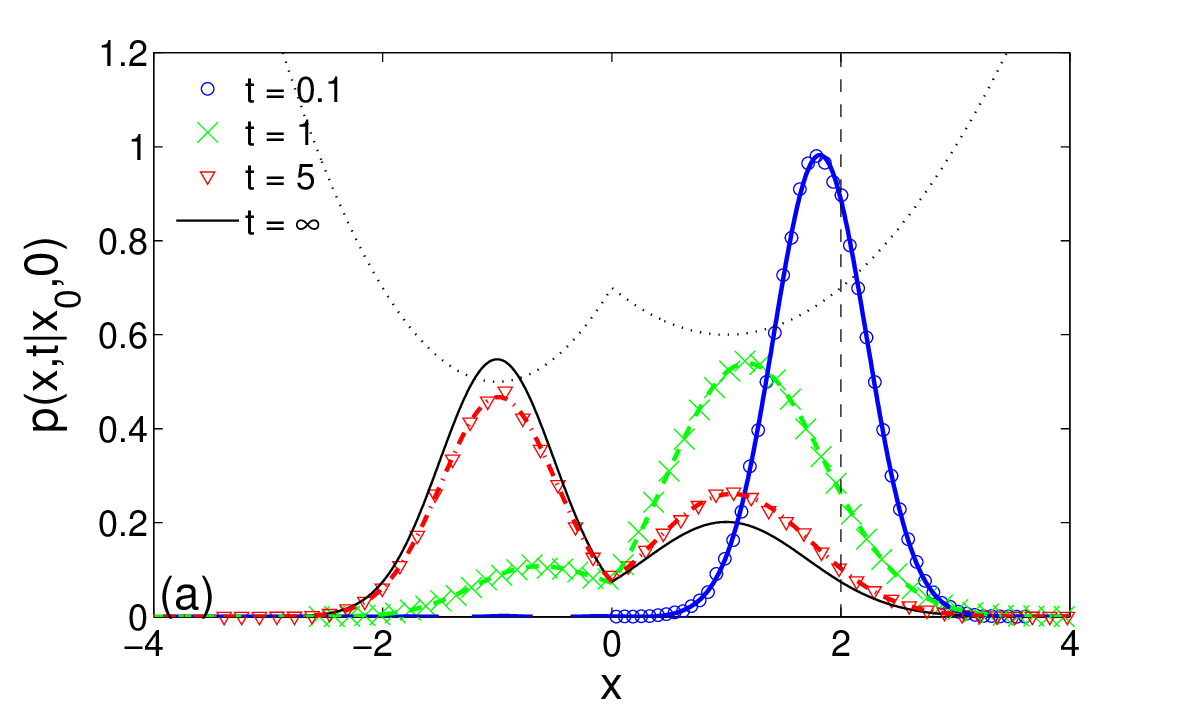}  % {pt_doublewell1.eps}
\includegraphics[width=70mm]{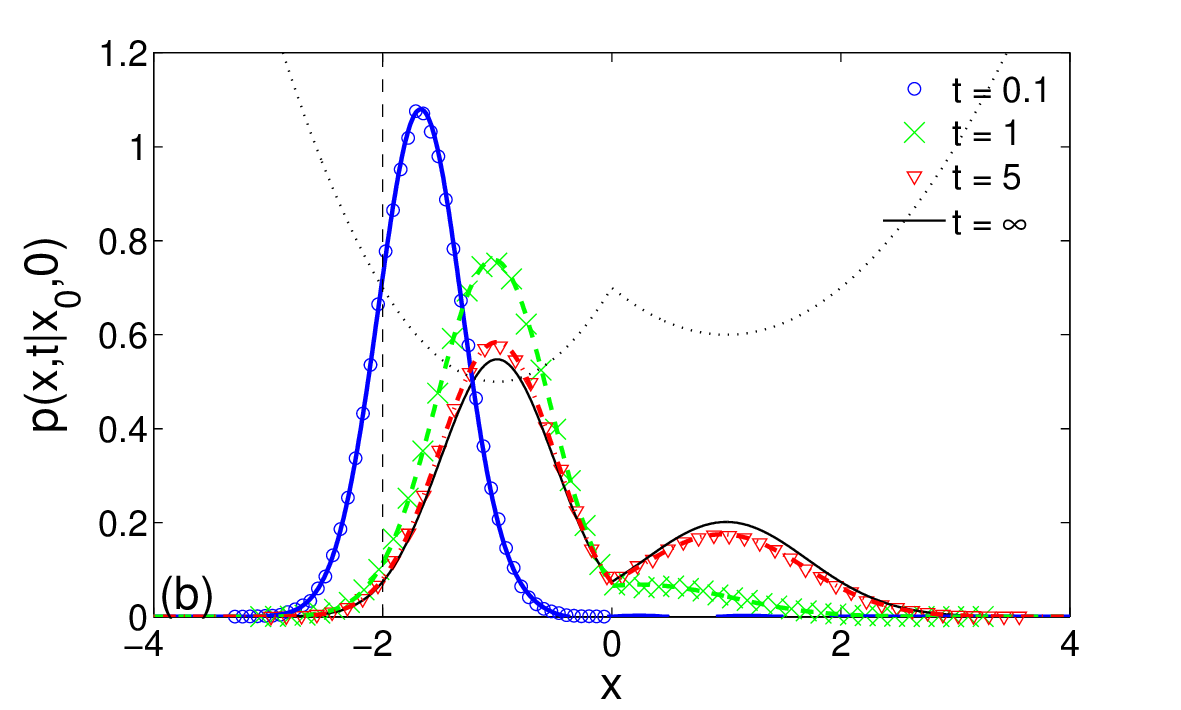}  % {pt_doublewell2.eps}
\end{center}
\caption{
Evolution of the probability density $p(x,t|x_0,t_0)$ for diffusion
under quadratic double-well potential (sketched by black dotted line)
with two minima at $\pm 1$ (i.e. $x_1 = x_2 = 1$), and $\kappa_1 = 2$,
$\kappa_2 = 1$.  Dashed vertical line indicates the starting point at
$t_0 = 0$: $x_0 = 2$ {\bf (a)} and $x_0 = -2$ {\bf (b)}.  Symbols
represent normalized histograms of arrival positions obtained by Monte
Carlo simulations of an adapted version of Eq. (\ref{eq:AR}) (with
time step $\delta = 10^{-3}$ and $10^5$ sample trajectories), while
lines show the spectral decomposition (\ref{eq:p_spectral}) with 50
terms.  We set $D = 1$.}
\label{fig:double_well}
% A_double_well_fig_pt();
\end{figure}

Figure \ref{fig:double_well} illustrates the evolution of the
probability density $p(x,t|x_0,t_0)$ for diffusion under quadratic
double-well potential with two minima at $\pm 1$ (i.e., $x_1 = x_2 =
1$) and two different dimensionless strengths: $\kappa_1 = 2$ and
$\kappa_2 = 1$.  One can see how the initial Dirac distribution,
concentrated at $x_0 = 2$ (Fig. \ref{fig:double_well}a) or $x_0 = -2$
(Fig. \ref{fig:double_well}b) and shown by dashed vertical line, is
progressively transformed into the equilibrium distribution $p_{\rm
eq}(x)$ (shown by black solid line).  Other diffusion characteristics
can be deduced from the probability density.

\subsection{Further extensions}
\label{sec:extensions}

The spectral approach is a general tool for computing FETs and other
first-passage quantities.  We briefly mention four straightforward
extensions.

(i) In one dimension, one can easily derive the splitting probability
$H(x_0)$, i.e., the probability to exit from one endpoint (e.g., $x_0
= L$) before the other ($x_0 = -L$).  The splitting probability is
governed by the stationary equation $\L^*_{x_0} H(x_0) = 0$ so that
$H(x_0)$ is given by a general solution in Eq. (\ref{eq:u_KummerF0})
with $\alpha = 0$.  Two constants $c_1$ and $c_2$ are set by boundary
conditions $H(L) = 1$ and $H(-L) = 0$, from which
\begin{equation}
H(x_0) = \frac{\erf(i\sqrt{\kappa}(x_0/L - \varphi)) + \erf(i\sqrt{\kappa}(1+\varphi))}
{\erf(i\sqrt{\kappa}(1-\varphi)) + \erf(i\sqrt{\kappa}(1+\varphi))} ,
\end{equation}
where we used $M(0,b,z) = 1$ and $M(1/2,3/2,z) = \frac{\sqrt{\pi}~
\erf(i\sqrt{z})}{2i \sqrt{z}}$.  Note that this expression can be
recognized in Eq. (\ref{eq:tmean_1d_gen}) for the mean exit time.

(ii) Dirichlet boundary conditions, $q(\pm L,t) = 0$, were imposed on
the FET probability density at both endpoints in order to stop the
process whenever it exits from the interval.  One can consider other
boundary value problems, e.g., with one reflecting endpoint or one/two
semi-reflecting points.  In this case, Dirichlet boundary condition at
one or both endpoints is replaced by Neumann or Robin boundary
conditions \cite{Gardiner}.  For instance, the condition
$\frac{\partial}{\partial x_0} q(x_0,t) = 0$ at $x_0 = -L$ describes
the reflecting barrier at $-L$.  The Robin boundary condition,
$\frac{\partial}{\partial x_0} q(x_0,t) + h q(x_0,t) = 0$, allows one
to consider partial absorptions/reflections for modeling various
transport mechanisms on the boundary and to switch continuously
between Neumann (pure reflections) and Dirichlet (pure absorptions)
cases by varying $h$ from $0$ to infinity
\cite{Collins49,Wilemski73,Sano79,Tachiya83,Sapoval94,Grebenkov06,Singer08}.
The solution can be obtained in the same way.

(iii) The first passage time to a single barrier can be deduced from
the first exit time from an interval by sending one endpoint to
infinity (see \ref{sec:FPT}).

(iv) A straightforward extension of the spectral approach allows one
to deduce FETs of continuous-time random walks (CTRW)
\cite{Metzler00,Metzler04}.  In this model, long stalling periods
between moves result in anomalous subdiffusion, when the mean-square
displacement evolves sublinearly with time: $\langle (X(t) -
X(0))^2\rangle \simeq 2D_\alpha t^\alpha$, with the exponent $0 <
\alpha < 1$ and the generalized diffusion coefficient $D_\alpha$.
The same derivations can be formally repeated for the fractional
Fokker-Planck equation that governs the survival probability of CTRWs.
In practice, it is sufficient to replace $s/D$ by $s^\alpha/D_\alpha$
in the Laplace domain that in time domain yields the replacement of
exponential functions $\exp(-\lambda_n t)$ by Mittag-Leffler functions
$E_\alpha(- \lambda_n D_\alpha t^\alpha/D)$ in spectral decompositions
such as Eq. (\ref{eq:S_spectral}) or similar.  As expected for CTRWs,
the mean exit time diverges due to long stalling periods while the
survival probability exhibits a power law decay $t^{-\alpha}$ at long
times instead of the exponential decay for normal diffusion.

\section*{Conclusion}
\addcontentsline{toc}{section}{Conclusion} 

We revised the classical problem of finding first exit times for
harmonically trapped particles.  Although the explicit formulas for
the moment-generating function $\langle e^{-s\tau}\rangle$ can be
found in standard textbooks (e.g., \cite{Borodin}), the computation of
the probability density and the survival probability through the
inverse Laplace transform requires substantial analysis of confluent
hypergeometric functions.  For didactic purpose, we reproduced the
main derivation steps and resulting spectral decompositions that
involve the eigenvalues and eigenfunctions of the governing
Fokker-Planck operator.  We also provided explicit formulas for the
mean exit time and discussed its asymptotic behavior in different
limits.  We considered the general case of non-centered harmonic
potential in one dimension (Ornstein-Uhlenbeck process with nonzero
mean) and the centered harmonic potential in higher dimensions (radial
Ornstein-Uhlenbeck process).  Both interior and exterior problems were
analyzed.

After revising this classical problem, we discussed some practical
issues.  First, we described a rapidly converging series
representation of confluent hypergeometric functions which is
particularly well suited for rapid numerical computation of
eigenvalues and eigenfunctions of the governing Fokker-Planck
operator.  Second, we showed how the mean exit time and the survival
probability can be used for the analysis of single-particle tracking
experiments with optically trapped tracers.  The derived formulas
allow one to choose the appropriate value of the optical tweezers'
stiffness and to detect in acquired trajectories the active periods
with nonzero force exerted by motor proteins.  Third, we mentioned the
relation of the first exit time problem to the dynamics of bond
dissociation under mechanical stress which plays an important role in
cell adhesion and motility.  Fourth, we considered an application of
FETs for algorithmic trading in stock markets in which buying or
selling signals are triggered when the difference between the current
and anticipated prices exceeds a prescribed threshold.  In a first
approximation, these events correspond to exits of an
Ornstein-Uhlenbeck process from an interval so that the FET statistics
can be used to estimate strategy holding periods and to choose the
appropriate threshold that ensures the desired transaction rate.
Fourth, we mentioned the relation to the distribution of first
crossing times of a moving boundary by Brownian motion.  Finally, we
discussed several extensions of the spectral approach, including
diffusion under quadratic double-well potential and anomalous
diffusion.

\section*{Acknowledgments}
The author acknowledges partial support by the grant
ANR-13-JSV5-0006-01 of the French National Research Agency.

\appendix
\addcontentsline{toc}{section}{Appendices} 
\addtocontents{toc}{\protect\setcounter{tocdepth}{-1}}

\section{Limit of large $\kappa$}
%\addcontentsline{toc}{section}{Appendix A: Limit of large $\kappa$} 

\subsection{Mean exit time}
\label{sec:tmean_large_kappa}

For large $\kappa$ or $\varphi$, Eq. (\ref{eq:tmean_1d_gen}) is not
appropriate for numerical computation of the mean exit time because
integrals and error functions are exponentially large.  Re-arranging
these terms, one can rewrite Eq. (\ref{eq:tmean_1d_gen}) as
\begin{eqnarray}
\label{eq:tmean_1d_gen2}
\langle \tau \rangle_{z_0} &=& \frac{L^2}{D} \frac{\sqrt{\pi}}{2\kappa} \Biggl\{ 
\frac{1 + e^{-2\kappa\varphi(1+z_0)-\kappa(1-z_0^2)} \frac{D(\sqrt{\kappa}(z_0-\varphi))}{D(\sqrt{\kappa}(1+\varphi))}}
{1 + e^{-4\kappa\varphi} \frac{D(\sqrt{\kappa}(1-\varphi))}{D(\sqrt{\kappa}(1+\varphi))}}  \\
\nonumber
&\times& \hspace*{-2mm} \int\limits_{\sqrt{\kappa}(\varphi-1)}^{\sqrt{\kappa}(1+\varphi)} \hspace*{-2mm} dz ~ e^{z^2} \erfc(z) 
- \int\limits_{\sqrt{\kappa}(\varphi-z_0)}^{\sqrt{\kappa}(1+\varphi)} \hspace*{-2mm} dz ~ e^{z^2} \erfc(z) \Biggr\} , 
\end{eqnarray}
where $z_0 = x_0/L$, and $D(x)$ is the Dawson function:
\begin{equation}
D(x) = e^{-x^2} \int\limits_0^x dt~ e^{t^2} ,
\end{equation}
which is related to the error function of imaginary argument as
\begin{equation}
\erf(ix) = \frac{2i}{\sqrt{\pi}} e^{x^2} D(x).
\end{equation}
For large $x$, the Dawson function decays as
\begin{equation}
D(x) \simeq \frac{1}{2x} + \frac{1}{4x^3} + \frac{3}{8x^5} + \ldots.  
\end{equation}

The relation (\ref{eq:tmean_1d_gen2}) allows one to compute the mean
exit time in the limit of large $\kappa$ and/or $\varphi$.  In fact,
since the Dawson function vanishes for large argument, the ratio in
front of the first integral in Eq. (\ref{eq:tmean_1d_gen2}) becomes
exponentially close to $1$ so that
\begin{equation}
\label{eq:tmean_1d_gen2_asympt}
\langle \tau \rangle_{z_0}  \simeq \frac{L^2}{D} \frac{\sqrt{\pi}}{2\kappa} 
\int\limits_{\sqrt{\kappa}(\varphi-1)}^{\sqrt{\kappa}(\varphi-z_0)} \hspace*{-2mm} dz ~ e^{z^2} \erfc(z)  . 
\end{equation}

Three situations can be considered separately.
\begin{enumerate}
\item
If $\varphi > 1$, the upper and lower limits of the above integral are
positive and large so that
\begin{equation}
\label{eq:tau_varphi_large}
\langle \tau \rangle_{z_0} \simeq \frac{L^2}{D} ~ \frac{1}{2\kappa} \ln \frac{\varphi - z_0}{\varphi-1}   \qquad (\kappa \gg 1),
\end{equation}
where we used the asymptotic relation
\begin{equation}
\int\limits_a^b dz ~ e^{z^2} \erfc(z) \simeq \frac{\ln(b/a)}{\sqrt{\pi}}  \qquad (a,b\gg 1).
\end{equation}
Note that Eq. (\ref{eq:tau_varphi_large}) is accurate already for
$\varphi \gtrsim 2$ (and $\kappa \geq 1$).

\item 
If $0 < \varphi < 1$ but $\kappa \to \infty$, the lower limit goes to
$-\infty$, and the integral exponentially diverges:
\begin{equation}
\label{eq:tmean_1d_gen_smallvarphi}
\langle \tau \rangle_0 \simeq \frac{L^2}{D} ~ \frac{\sqrt{\pi} ~ e^{\kappa(1-\varphi)^2}}{2\kappa^{3/2}(1-\varphi)}  \qquad (\kappa \gg 1)
\end{equation}
(here the starting point is set to $0$, but the result holds for all
$z_0$ not too close to $1$).  This relation is valid for any $0 <
\varphi < 1$.  Setting formally $\varphi = 0$, one gets the relation
which is twice larger than the asymptotic Eq. (\ref{eq:tmean_1d})
derived for $\varphi = 0$.  The missing factor $2$ can be retrieved
from the ratio in front of the first integral in
Eq. (\ref{eq:tmean_1d_gen2}).  The difference between the cases
$\varphi = 0$ and $\varphi > 0$ (small but strictly positive) can also
be explained by the following argument.  For non-symmetric case
($\varphi > 0$), the right endpoint $x_0 = L$ is closer to the minimum
position $\hat{x}$ than the left endpoint $x_0 = -L$.  When $\kappa$
is large, the probability of large deviations from $\hat{x}$ rapidly
decays with the distance so that the probability of exiting through
the left endpoint is exponentially smaller than that from the right
endpoint.  In other words, the above relation essentially describes
the mean exit time from the right endpoint.  In turn, when $\varphi =
0$ (and thus $\hat{x} = 0$), both endpoints are equivalent that
doubles the chances to exit and thus twice reduces the mean exit time.

\item
In the marginal case $\varphi = 1$, the integral in
Eq. (\ref{eq:tmean_1d_gen2_asympt}) grows logarithmically with
$\kappa$.  One can split the integral by an intermediate point
$\bar{z} \gg 1$ so that
\begin{equation*}
\int\limits_0^{\bar{z}} dz~ e^{z^2}~ \erfc(z) + \int\limits_{\bar{z}}^{\sqrt{\kappa} (1 - z_0)} dz~ e^{z^2}~ \erfc(z) \simeq 
\frac{1}{\sqrt{\pi}} \ln \frac{\sqrt{\kappa} (1 - z_0)}{c(\bar{z})} ,
\end{equation*}
where 
\begin{equation*}
c(\bar{z}) \equiv \bar{z} \exp\left(-\sqrt{\pi} \int\limits_0^{\bar{z}} dz~ e^{z^2}~ \erfc(z)\right) 
\longrightarrow 0.375\ldots \quad (\bar{z}\to\infty).
\end{equation*}
We get therefore
\begin{equation}
\label{eq:tau_varphi1}
\langle \tau \rangle_{z_0} \simeq \frac{L^2}{D} ~ \frac{1}{2\kappa} \ln \frac{\sqrt{\kappa}(1 - z_0)}{0.375\ldots}   \qquad (\kappa \gg 1).
\end{equation}
This asymptotic relation is accurate starting from $\sqrt{\kappa}
(1 - z_0) \gtrsim 2$.

\end{enumerate}

\subsection{Eigenvalues (interior problem)}
\label{sec:large_kappa}

For large $\kappa$, we search for positive solutions $\alpha_n$ of the
equation $M\bigl(-\frac{\alpha_n^2}{4\kappa}, b, \kappa\bigr) = 0$ in
the form: $\alpha_n^2/(4\kappa) = n - \ve$, where $\ve$ is a small
parameter, and $n = 0,1,2,\ldots$ One gets then
\begin{eqnarray*}
0 = M\left(-n + \ve; b; \kappa\right) 
&\simeq& \sum\limits_{j=0}^n \frac{(-n)(-n+1)\ldots (-n+j-1) \kappa^j}{b^{(j)}~ j!}  \\
&+& \ve (-1)^n n! \sum\limits_{j=n+1}^\infty \frac{(-n+j-1)! ~\kappa^j}{b^{(j)}~ j!} + O(\ve^2)  ,
\end{eqnarray*}
from which the small parameter $\ve$ can be determined as
\begin{equation}
\ve \simeq - \frac{S_1}{(-1)^n n! S_2} ,
\end{equation}
where $S_1$ and $S_2$ denote two above sums.  The second sum can be
written as
\begin{equation*}
\fl
S_2 = \sum\limits_{j=n+1}^\infty \frac{(-n+j-1)! ~\kappa^j}{b^{(j)}~ j!} = 
\Gamma(b) \sum\limits_{j=0}^\infty \frac{\kappa^{j+n+1}}{\Gamma(b+j+1+n) (j+1)\ldots (j+1+n)} .
\end{equation*}
This expression can be obtained by integrating $n+1$ times the
Mittag-Leffler function $E_{1,b+n+1}(\kappa)$ which asymptotically
behaves as $E_{1,b+n+1}(\kappa) \simeq \kappa^{-b-n} e^{\kappa} (1 +
O(1/\kappa))$ as $\kappa \gg 1$.  Since the integration does not
change the leading term, one concludes that
\begin{equation*}
S_2 \simeq \Gamma(b) \kappa^{-b-n} e^{\kappa} (1 + O(1/\kappa))  \qquad (\kappa \gg 1).
\end{equation*}
Keeping the highest-order term in the first sum, $S_1 \simeq (-1)^n
\kappa^n/b^{(n)}$, one gets
\begin{equation*} 
\ve \simeq - \frac{\kappa^{b+2n}}{n! \Gamma(b+n)} e^{-\kappa}  ,
\end{equation*} 
from which we obtain the asymptotic behavior of the positive solution
$\alpha_n$ as $\kappa \gg 1$:
\begin{equation}
\alpha^2_n \simeq 4\kappa \biggl[n + \frac{\kappa^{b+2n} e^{-\kappa}}{n! \Gamma(b+n)} \biggr]    \qquad (n = 0,1,2,\ldots) .
\end{equation}
In particular, the smallest solution $\alpha_0$ exponentially decays
with $\kappa$,
\begin{equation}
\label{eq:alpha1_d}
\alpha_0^2 \simeq \frac{4\kappa^{1+b}}{\Gamma(b)}~ e^{-\kappa} \qquad (\kappa \gg 1),
\end{equation}
while the other eigenvalues grow linearly with $\kappa$:
\begin{equation}
\label{eq:alphan_b}
\alpha_n^2 \simeq 4\kappa n \qquad (\kappa \gg 1,~ n=1,2,\ldots),
\end{equation}
and the first-order correction $\ve$ decays exponentially fast.  This
asymptotic behavior can be related to equidistant energy levels of a
quantum harmonic oscillator (see \ref{sec:QH}).

Since $\alpha_0$ rapidly vanishes, the first eigenfunction approaches
the unity: $M\bigl(- \frac{\alpha_0^2}{4\kappa}, \frac{d}{2}, \kappa
z^2 \bigr) \to 1$.  As a consequence, the normalization constant is
simply $\beta_0^2 \approx 2\kappa^{d/2}/\Gamma(d/2)$ so that
$w_0\simeq 1$, because $M(1,b,z) = \Gamma(b) E_{1,b}(z)$.

\subsection{Eigenvalues (exterior problem)}
\label{sec:large_kappa_ext}

For the exterior problem, we consider the asymptotic behavior of
solutions of $U\bigl(-\frac{\alpha^2}{4\kappa}, b, \kappa) = 0$ as
$\kappa \to 0$.  For non-integer $b$, one can use
Eq. (\ref{eq:KummerU}) to write in the lowest order in $\kappa$
\begin{equation}
0 = U\biggl(-\frac{\alpha^2}{4\kappa}, b, \kappa \biggr) \simeq \frac{\Gamma(1-b)}{\Gamma(1-b- \frac{\alpha^2}{4\kappa})} 
+ \frac{\Gamma(b-1)}{\Gamma(-\frac{\alpha^2}{4\kappa})} \kappa^{1-b}  .
\end{equation}
For $b < 1$, $\kappa^{1-b}$ is a small parameter so that the first
term has to be small.  Setting $1-b-\frac{\alpha^2}{4\kappa} = -n +
\ve$ (with $n = 0,1,2,\ldots$) one gets
\begin{equation}
\ve = (-1)^{n-1} \frac{\Gamma(b-1)}{n!~ \Gamma(b-1-n) \Gamma(1-b)} ~ \kappa^{1-b} ,
\end{equation}
from which
\begin{equation}
\alpha_n^2 \simeq 4\kappa\biggl(1-b + n + \frac{(-1)^n \Gamma(b-1)}{n!~ \Gamma(b-1-n) \Gamma(1-b)} ~ \kappa^{1-b} + \ldots\biggr) .
\end{equation}
In turn, if $b > 1$, $\kappa^{1-b}$ is a large parameter so that the
second term has to be small.  Setting $-\frac{\alpha^2}{4\kappa} = -n
+ \ve$, one gets
\begin{equation}
\ve = (-1)^{n-1} \frac{\Gamma(1-b)}{n!~ \Gamma(1-b-n) \Gamma(b-1)} ~ \kappa^{b-1} ,
\end{equation}
from which
\begin{equation}
\label{eq:alphan_D_ext}
\alpha_n^2 \simeq 4\kappa\biggl(n + \frac{(-1)^n \Gamma(1-b)}{n!~ \Gamma(1-b-n) \Gamma(b-1)} ~ \kappa^{b-1} + \ldots\biggr)  .
\end{equation}

For integer $b$, the analysis is more subtle and is beyond the scope
of this paper.  We just checked numerically that $\alpha_0^2 \propto
\kappa^b$ as $\kappa\to 0$ for $b = 1$ and $b = 2$ that corresponds to
dimensions $d = 2$ and $d = 4$.

\section{Confluent hypergeometric functions}  
\label{sec:Kummer}

For the sake of completeness, we summarize selected relations between
special functions that are often used to describe first passage times
of Ornstein-Uhlenbeck processes (see \cite{Abramowitz} for details).
After that, we describe a rapidly converging representation of
confluent hypergeometric functions.

\subsection{Relations}
\label{sec:relations}

The Kummer confluent hypergeometric function $M(a,b,z) =~
_1F_1(a;b;z)$ defined in Eq. (\ref{eq:KummerM}), satisfies the
Kummer's equation:
\begin{equation}
z y'' + (b-z) y' - a y = 0 .
\end{equation}
For $b = 1/2$, this equation is also related to the Weber's equation
\begin{equation}
\label{eq:Weber}
y'' - (z^2/4 + c) y = 0 ,
\end{equation}
which has two independent solutions: $e^{-z^2/4} M(c/2 + 1/4, 1/2,
z^2/2)$ (even) and $z e^{-z^2/4} M(c/2 + 3/4, 3/2, z^2/2)$ (odd).
These solutions are often expressed through the parabolic cylinder
function $D_\nu(z)$, which satisfies Eq. (\ref{eq:Weber}) with $\nu =
-c-1/2$:
\begin{eqnarray}
\label{eq:CylinderD1}
D_{\nu}(z) &=& \frac{\cos(\frac{\pi \nu}{2}) \Gamma(\frac{1+\nu}{2})}{\sqrt{\pi}~ 2^{-\nu/2}}~ 
e^{-z^2/4} ~ M\left(-\frac{\nu}{2}, \frac12, \frac{z^2}{2}\right) \\
\nonumber
&+& \frac{\sin(\frac{\pi \nu}{2}) \Gamma(\frac{2+\nu}{2})}{\sqrt{\pi}~ 2^{-(\nu+1)/2}}~ 
e^{-z^2/4} ~ z M\left(-\frac{\nu}{2}+\frac12, \frac32, \frac{z^2}{2}\right) \\
\label{eq:CylinderD2}
&=& 2^{\nu/2} e^{-z^2/4} U\left(-\frac{\nu}{2}, \frac12, \frac{z^2}{2}\right) 
\end{eqnarray}
(the last relation is valid only for $\Re\{z\} \geq 0$).

The confluent hypergeometric functions $M(a,b,z)$ and $U(a,b,z)$ are
also related to the Whittaker functions $M_{a,b}(z)$ and $W_{a,b}(z)$
\cite{Abramowitz}
\begin{eqnarray*}
M_{a,b}(z) &=& e^{-z/2} z^{b+1/2} ~ M(1/2 + b -a , 1+2b, z) , \\
W_{a,b}(z) &=& e^{-z/2} z^{b+1/2} ~ U(1/2 + b -a , 1+2b, z) . 
\end{eqnarray*}

The following relations help to analyze the Brownian motion limit
\cite{Borodin}
\begin{eqnarray}
&& \lim\limits_{\kappa\to 0} M\left(\frac{a}{4\kappa},b+1,\kappa x\right) = 2^b \Gamma(b+1) \frac{I_b(\sqrt{xa})}{(xa)^{b/2}} , \\
&& \lim\limits_{\kappa\to 0} \kappa^b \Gamma\left(\frac{a}{4\kappa}\right) U\left(\frac{a}{4\kappa},b+1,\kappa x\right) 
 = 2^{1-b} \frac{K_b(\sqrt{xa})}{(x/a)^{b/2}} ,
\end{eqnarray}
where $I_\nu(z)$ and $K_\nu(z)$ are the modified Bessel functions of
the first and second kind, respectively.

The asymptotic expansions for large $|z|$ (and fixed $a$ and $b$) are
\cite{Abramowitz} (Sec. 13.5):
\begin{eqnarray}
\label{eq:M_asympt}
\fl
M(a,b,z) &\simeq& \frac{e^{z} z^{a-b} \Gamma(b)}{\Gamma(a)} \left(\sum\limits_{n=0}^{N_1-1} \frac{(b-a)^{(n)} (1-a)^{(n)}}{n!~ z^n} + O(|z|^{-N_1})\right) \\
\nonumber
\fl
&+& \frac{e^{\pm \pi i a} z^{-a} \Gamma(b)}{\Gamma(b-a)} \left(\sum\limits_{n=0}^{N_2-1} \frac{(a)^{(n)} (1+a-b)^{(n)}}{n!~ (-z)^{n}} + O(|z|^{-N_2})\right), \\
\fl
\label{eq:U_asympt}
U(a,b,z) &\simeq& z^{-a} \left(\sum\limits_{n=0}^{N-1} \frac{(a)^{(n)} (1+a-b)^{(n)}}{n!~ (-z)^{n}} + O(|z|^{-N})\right) \quad (|\arg(z)| < 3\pi/2),
\end{eqnarray}
where the upper [resp., lower] sign in the second line is taken if
$-\pi/2 < \arg(z) < 3\pi/2$ [resp., $-3\pi/2 < \arg(z) \leq -\pi/2$],
and $N$, $N_1$, and $N_2$ are truncation orders.

\subsection{Computation}
\label{sec:computation}

\subsubsection*{Series representations.}

The computation of the Kummer function $M(a,b,z)$ by direct series
summation in Eq. (\ref{eq:KummerM}) is not convenient for large $|a|$.
For this case, two equivalent representations were proposed:
\begin{enumerate}
\item
\begin{equation}
M(a,b,z) = \Gamma(b) e^{z/2} 2^{b-1} \sum\limits_{n=0}^\infty A_n z^n \frac{J_{b-1+n}\bigl(\sqrt{z(2b-4a)}\bigr)}{\bigl(\sqrt{z(2b-4a)}\bigr)^{b-1+n}} ,
\end{equation}
where the coefficients $A_n$ are defined by
\begin{equation*}
\fl
A_0 = 1, \quad A_1 = 0, \quad A_2 = b/2, \quad nA_n = (n-2+b)A_{n-2} + (2a-b) A_{n-3}
\end{equation*}
(see \cite{Abramowitz}, Sec. 13.3.7, and \cite{Luke}, Sec. 4.8).  Note
that the coefficients $A_n$ depend on $a$ and grow with $|a|$.

\item
\begin{equation}
\label{eq:KummerM_as}
M(a,b,z) = \Gamma(b) e^{z/2} 2^{b-1} \sum\limits_{n=0}^\infty p_n(b,z) \frac{J_{b-1+n}(\sqrt{z(2b-4a)})}{(\sqrt{z(2b-4a)})^{b-1+n}} ,
\end{equation}
where $p_n(b,z)$ are the Buchholz polynomials in $b$ and $z$ (see
\cite{Buchholz}, Sec. 7.4).  These polynomials are less explicit than
the coefficients $A_n$, but they are independent of $a$.  As a
consequence, this representation is particularly convenient for large
$|a|$.
\end{enumerate}

The recurrence relations for the Buchholz polynomials were derived in
\cite{Abad95}:
\begin{equation}
p_n(b,z) = \frac{(iz)^n}{n!} \sum\limits_{k=0}^{[n/2]}  \binom{n}{2k} f_k(b) g_{n-2k}(z) , 
\end{equation}
where the polynomials $f_k(b)$ and $g_k(z)$ are defined recursively by
\begin{eqnarray}
f_k(b) &=& - \left(\frac{b}{2} - 1\right) \sum\limits_{j=0}^{k-1} \binom{2k-1}{2j} \frac{4^{k-j} |B_{2(k-j)}|}{k-j} ~ f_j(b),  \quad f_0(b) = 1, \\  
g_k(z) &=& - \frac{iz}{4} \sum\limits_{j=0}^{[(k-1)/2]} \binom{k-1}{2j} \frac{4^{j+1} |B_{2(j+1)}|}{j+1} ~ g_{k-2j-1}(z), \quad g_0(z) = 1 , 
\end{eqnarray}
and $B_{2j}$ are the Bernoulli numbers.  Using the recurrence
relations between Bessel functions, $\frac{2\nu}{x} J_\nu(x) =
J_{\nu-1}(x) + J_{\nu+1}(x)$, one can express
\begin{equation*}
J_{b-1+n}(x) = P_n(1/x) J_{b-1}(x) + Q_n(1/x) J_b(x),
\end{equation*}
where the polynomials $P_n(z)$ and $Q_n(z)$ are defined recursively
\begin{eqnarray*}
&& P_0(z) = 1, \quad P_1(z) = 0, \quad P_{n+1}(z) = 2(b-1 + n) z P_n(z) - P_{n-1}(z) , \\
&& Q_0(z) = 0, \quad Q_1(z) = 1, \quad Q_{n+1}(z) = 2(b-1 + n) z Q_n(z) - Q_{n-1}(z) .
\end{eqnarray*}
We get therefore the following expansion which rapidly converges for
large $x$ and moderate $z$:
\begin{equation}
\label{eq:KummerM_expansion}
M(a,b,z) =  e^{z/2}  \sum\limits_{n=0}^\infty p_n(b,z) \left[F_b(x) \frac{P_n(1/x)}{x^n} + G_b(x) \frac{Q_n(1/x)}{x^{n-1}} \right] , 
\end{equation}
where $x = \sqrt{z(2b-4a)}$, and
\begin{equation}
F_b(x) = \Gamma(b) 2^{b-1} x^{1-b} J_{b-1}(x) , \qquad  G_b(x) = \Gamma(b) 2^{b-1} x^{-b} J_{b}(x) .
\end{equation}
In particular, for $b = d/2$, one has
\begin{equation}
\begin{array}{c | c | c}  d & F_b(x) & G_b(x) \\  \hline
1 & \cos(x) & \sin(x)/x \\ 
2 &  J_0(x) & J_1(x)/x  \\ 
3 & \sin(x)/x & (\sin(x) - x\cos(x))/x^3 \\  \end{array}
\end{equation}
The above recursive relations allow one to compute rapidly the
polynomials $p_n(b,z)$, $P_n(1/x)$ and $Q_n(1/x)$.  The series
(\ref{eq:KummerM_expansion}) can be truncated after 5-10 terms when
$|az|$ is large enough, and $z$ is not too large (see \cite{Abad95}
for several examples).  

According to Eq. (\ref{eq:KummerU}), one can apply this method to
compute the Tricomi confluent hypergeometric function $U(a,b,z)$ for
non-integer $b$.  Other series expansions for $U(a,b,z)$ are discussed
in \cite{Abad97,Temme83}.  For integer $b$, one can substitute $b_\ve
= b+\ve$ into Eq. (\ref{eq:KummerU}) and then take the limit $\ve \to
0$.  This extension by continuity yields \cite{Abramowitz}
\begin{eqnarray*}
\fl
U(a,b,z) &=& \frac{(-1)^b}{(b-1)!~\Gamma(a-b+1)} \Biggl\{M(a,b,z) \ln z  
+ \frac{(b-2)!}{\Gamma(a)} \sum\limits_{k=0}^{b-2} \frac{(a-b+1)^{(k)}  z^{k-b+1}}{k!~(2-b)^{(k)}}  \\
\fl &+& \sum\limits_{k=0}^\infty \frac{a^{(k)} z^k}{k!~ b^{(k)}} \biggl(\psi(a+k) - \psi(1+k) - \psi(b+k)\biggr)  \Biggr\} ,
\qquad b = 1,2,\ldots,
\end{eqnarray*}
where $\psi(z) = \Gamma'(z)/\Gamma(z)$ is the digamma function, and
the intermediate sum is omitted for $b = 1$.  In practice, one can
apply the above numerical scheme to rapidly compute $U(a,b_\ve,z)$
through $M(a,b_\ve,z)$ with several non-integer $b_\ve$ approaching
the integer $b$, and then to extrapolate them in the limit $b_\ve \to
b$.

Taking the derivative of Eq. (\ref{eq:KummerM_as}) with respect to $a$
and using the relation $J'_\nu(x) = \frac{\nu}{x} J_\nu(x) -
J_{\nu+1}(x)$, one obtains
\begin{equation}
\label{eq:dKummerM_as}
\frac{\partial}{\partial a} M(a,b,z) = \Gamma(b) e^{z/2} 2^{b} z
\sum\limits_{n=0}^\infty p_n(b,z) \frac{J_{b+n}(\sqrt{z(2b-4a)})}{(\sqrt{z(2b-4a)})^{b+n}} 
\end{equation}
or, equivalently,
\begin{equation}
\label{eq:dKummerM_as2}
\fl 
\frac{\partial}{\partial a} M(a,b,z) = 2z e^{z/2} \sum\limits_{n=0}^\infty p_n(b,z) 
\left[F_b(x) \frac{P_{n+1}(1/x)}{x^{n+1}} + G_b(x) \frac{Q_{n+1}(1/x)}{x^n} \right] ,
\end{equation}
with $x = \sqrt{z(2b-4a)}$.  This expression allows one to rapidly
compute the coefficients $w_n$ in the spectral representation of the
survival probability.  Similar relation can be derived for
$\frac{\partial}{\partial a} U(a,b,z)$ using Eq. (\ref{eq:KummerU})
for non-integer $b$.  Finally, one can also apply these formulas for
computing the parabolic cylinder function $D_\nu(z)$ and its
derivative $\frac{\partial}{\partial \nu} D_\nu(z)$ which are used to
characterize the first passage time to a single barrier
(\ref{sec:FPT}).

\subsubsection*{Integral representations.}

The above scheme is convenient for large $|a|$ and moderate $|z|$.
However, if $|a|$ is moderate while $|z|$ is large, the numerical
convergence of the above series is slowed down due to a rapid growth
of Buchholz polynomials with $z$.  In addition, the computation of the
Tricomi function $U(a,b,z)$ as a linear combination (\ref{eq:KummerU})
of two large Kummer functions can result in significant round-off
errors at large $z$.  In this case, one can apply a different
technique which relies on integral representations of confluent
hypergeometric functions.

For the Kummer function $M(a,b,z)$, one can use the following integral
representation for $\Re\{b-a\} > 0$ 
\footnote{See \url{http://dlmf.nist.gov/13.16.E3}}
\begin{equation}
M(a,b,z) = \frac{e^{z} z^{\frac{1-b}{2}} \Gamma(b)}{\Gamma(b-a)} \int\limits_0^\infty dt~ e^{-t}~ t^{\frac{b-1}{2}-a}~ J_{b-1}(2\sqrt{zt}) .
\end{equation}
This representation is convenient for computing eigenvalues and
eigenfunctions because $a = - \alpha^2/(4\kappa) <0$ and $b = d/2 >
0$.

The Tricomi function $U(a,b,z)$ has an integral representation for
positive $a$ \cite{Abramowitz}
\begin{equation}
\label{eq:U_int}
U(a,b,z) = \frac{1}{\Gamma(a)} \int\limits_0^\infty dt~ e^{-zt}~ t^{a-1}~ (1+t)^{b-a-1}  \quad (\Re\{a\} >0, \Re\{z\} > 0).
\end{equation}
When $a$ is negative, one can use the recurrence relation to increase
$a$:
\begin{equation}
\fl
U(a-1,b,z) + (b-2a-z) U(a,b,z) + a(a+1-b) U(a+1,b,z) = 0.
\end{equation}
Applying this relation repeatedly, one gets
\begin{equation}
U(a,b,z) = p_n(a,b,z) U(a+n,b,z) + q_n(a,b,z) U(a+n+1,b,z) ,
\end{equation}
where the polynomials $p_n(a,b,z)$ and $q_n(a,b,z)$ can be rapidly
computed through recurrence relations:
\begin{eqnarray*}
p_n(a,b,z) &=& q_{n-1}(a,b,z) - (b - 2(a+n) - z) p_{n-1}(a,b,z) , \qquad p_0 = 1, \\
q_n(a,b,z) &=& - (a+n) (a+n+1-b) p_{n-1}(a,b,z) , \qquad \qquad q_0 = 0 .
\end{eqnarray*}
Choosing $n$ such that $a+n > 0$, one can express $U(a,b,z)$ in terms
of $U(a+n,b,z)$ and $U(a+n+1,b,z)$ which are found by numerical
integration of Eq. (\ref{eq:U_int}).  If $z$ is too large, it is
convenient to divide each recurrence relation by $z$ and to consider
them as polynomials of $1/z$.  The resulting value can be compared
with the asymptotic expansion (\ref{eq:U_asympt}).

\section{First passage time to a single barrier}
\label{sec:FPT}

The first passage times (one-barrier problem) for harmonically trapped
particles have attracted more attention than the first exit times
(two-barrier problem)
\cite{Siegert51,Ricciardi88,Leblanc00,Going03,Alili05,Yi10}.  In
general, the first passage time $\tau_\ell$ to a single barrier at
$\ell > 0$ in one dimension can be found following the steps from
Sec. \ref{sec:spectral_1d}.  In practice, these results can be deduced
from the FET statistics.  If the starting point $x_0$ lies on the
right to $\ell$ (i.e., $x_0 > \ell$), this problem is equivalent to
the exterior problem to reach the interval $[-\ell,\ell]$ from outside
(see Sec. \ref{sec:exterior}).  In turn, if $0 < x_0 < \ell$, the FPT
to a single barrier $\ell$ can be deduced from the FET from the
interval $[-a,\ell]$ in the limit $a\to
\infty$.

In order to illustrate this point, we focus on the moment-generating
function $\tilde{q}(x_0,s)$ given by Eq. (\ref{eq:char_1d}), with
$\ell = L(1-\varphi)$ and $a = L(1+\varphi)$.  Setting $1-\varphi =
\ve$, we consider the limit $\ve \to 0$, for which $L = \ell/\ve
\to\infty$ and $1-\varphi = \ve \to 0$ so that $a\to\infty$ while
$\ell$ is kept fixed.  The asymptotic behavior of functions
$m_{\alpha,\kappa}^{(1,2)}$ from Eq. (\ref{eq:mm}) as $\ve \to 0$ can
be easily found:
\begin{eqnarray*}
m_{\alpha,\kappa}^{(1)}(1-\varphi) &\simeq&  M\biggl(a ,\frac12, y\biggr), \\
m_{\alpha,\kappa}^{(2)}(1-\varphi) &\simeq&  \ve M\biggl(a+\frac12,\frac32, y\biggr), \\
m_{\alpha,\kappa}^{(1)}(-1-\varphi) &\simeq&  \frac{\Gamma(1/2)}{\Gamma(a)} (4y/\ve^2)^{a-1/2} e^{4y/\ve^2}, \\
m_{\alpha,\kappa}^{(2)}(-1-\varphi) &\simeq&  -2 \frac{\Gamma(3/2)}{\Gamma(a+1/2)} (4y/\ve^2)^{a-1} e^{4y/\ve^2}, 
\end{eqnarray*}
where we replaced $\alpha$ and $\kappa$ by $-Ds/L^2$ and
$kL^2/(2D\gamma)$, introduced short notations $a = s\gamma/(2k)$ and
$y = k\ell^2/(2D\gamma)$, and used the asymptotic relation
(\ref{eq:M_asympt}) for the last two functions.  Substituting the
above expressions into Eq. (\ref{eq:char_1d}), one deduces in the
limit $\ve \to 0$
\begin{equation}
\tilde{q}(x_0,s) = \frac{M\bigl(a ,\frac12, y_0\bigr) + 2 \frac{\Gamma(a+1/2)}{\Gamma(a)} \sqrt{y_0}M\bigl(a+1/2 ,\frac32, y_0\bigr)}
{M\bigl(a ,\frac12, y\bigr) + 2 \frac{\Gamma(a+1/2)}{\Gamma(a)} \sqrt{y} M\bigl(a+1/2 ,\frac32, y\bigr)} ,
\end{equation}
where $y_0 = kx_0^2/(2D\gamma)$.  Using Eq. (\ref{eq:CylinderD1}), one
can alternatively write the moment-generating function as
\begin{equation}
\label{eq:char_1d_FPT}
\tilde{q}(x_0,s) = \exp\left(\frac{k(x_0^2-\ell^2)}{4D\gamma}\right) \frac{D_{-s\gamma/k}\biggl(-x_0\sqrt{\frac{k}{D\gamma}}\biggr)}
{D_{-s\gamma/k}\biggl(-\ell\sqrt{\frac{k}{D\gamma}}\biggr)}  \qquad
(0 \leq x_0 \leq \ell).
\end{equation}
Note also that Eq. (\ref{eq:char_D_ext}) for the exterior case $x_0 >
\ell$ can also be written in terms of the parabolic cylinder function
$D_\nu(z)$ according to Eq. (\ref{eq:CylinderD2}):
\begin{equation}
\tilde{q}(x_0,s) = \exp\left(\frac{k(x_0^2-\ell^2)}{4D\gamma}\right) \frac{D_{-s\gamma/k}\biggl(x_0\sqrt{\frac{k}{D\gamma}}\biggr)}
{D_{-s\gamma/k}\biggl(\ell\sqrt{\frac{k}{D\gamma}}\biggr)}  \qquad (x_0 > \ell), 
\end{equation}
in agreement with \cite{Borodin} (see also
\cite{Jeanblanc,Ricciardi88}).

The inverse Laplace transform yields the probability density
$p_{x,a}(t)$ \cite{Jeanblanc,Alili05}
\begin{equation}
q(x_0,t) = - \frac{k}{\gamma} \exp\left(\frac{k(x_0^2 - \ell^2)}{4D\gamma}\right)
\sum\limits_{n=1}^\infty \frac{D_{\nu_n}\biggl(\pm x_0 \sqrt{\frac{k}{D\gamma}}\biggr)}{D'_{\nu_n}\biggl(\pm \ell\sqrt{\frac{k}{D\gamma}}\biggr)} 
e^{-\nu_n kt/\gamma} ,
\end{equation}
where $0 < \nu_1 < ... < \nu_n < ...$ are the zeros of the function
$D_\nu(\pm \ell\sqrt{k/(D\gamma)})$, and $D'_{\nu_n}(z)$ is the
derivative of $D_\nu(z)$ with respect to $\nu$, evaluated at point
$\nu = \nu_n$ \cite{Jeanblanc} (p. 154).  The signs plus and minus
correspond to $x_0 > \ell$ and $x_0 < \ell$, respectively.  Both
$D_\nu(z)$ and $D'_\nu(z)$ can be rapidly evaluated by the numerical
scheme presented in \ref{sec:computation}.

In the special case $\ell = 0$, the FET probability density gets a
simple explicit form:
\begin{equation}
\label{eq:q_ell0}
q(x_0,t) = \frac{x_0}{\sqrt{4\pi D}} \left(\frac{k/\gamma}{\sinh (kt/\gamma)}\right)^{3/2}  
 \exp\left(- \frac{k x_0^2}{4D\gamma} \frac{e^{- kt/\gamma }}{\sinh (kt/\gamma)} + \frac{kt}{2\gamma}\right) . 
\end{equation}
In the limit $k\to 0$, one retrieves the classical formula for the FPT
of Brownian motion at the origin
\begin{equation}
q(x_0,t) = \frac{x_0}{\sqrt{4\pi D t^3}} \exp\left(- \frac{x_0^2}{4D t}\right)  \qquad (k=0).
\end{equation}

\section{Quantum harmonic oscillator}
\label{sec:QH}

The eigenvalue problem (\ref{eq:u_diffeq}) with $b = 1/2$ is closely
related to eigenstates of a quantum harmonic oscillator of mass $m$
and frequency $\omega$ \cite{Merzbacher}.  In fact, the eigenstates
$\psi_n$ and energies $E_n$ of the Hamiltonian $H =
\frac{\hat{p}^2}{2m} + \frac{m\omega^2 x^2}{2m}$ satisfy the
time-independent Schr\"odinger equation
\begin{equation}
\biggl[- \frac{\hbar^2}{2m} \partial_x^2 + \frac{m\omega^2 x^2}{2}\biggr] \psi(x) = E \psi(x),
\end{equation}
where $\hat{p} = -i\hbar \partial_x$ is the momentum operator, and
$\hbar$ is the (reduced) Planck constant.  In terms of the
dimensionless coordinate $z = x \sqrt{2m\omega/\hbar}$, the above
Schr\"odinger equation is reduced to the Weber's equation
(\ref{eq:Weber}), with $c = -\frac{E}{\hbar \omega}$.  Setting
$\psi(z) = e^{-z^2/4} \tilde{u}(z)$ yields $\tilde{u}'' - z \tilde{u}'
- (c+1/2)\tilde{u} = 0$, from which the rescaling $u(x) =
\tilde{u}\bigl(\sqrt{k/(D\gamma)} (x-\hat{x})\bigr)$ implies
Eq. (\ref{eq:u_diffeq}), with $\lambda = - \frac{k}{\gamma}(c+1/2)$.
As a consequence, the energies of the quantum oscillator and the
eigenvalues of the FP operator are simply related as: $\lambda =
\frac{k}{\gamma} (\frac{E}{\hbar\omega} - \frac12)$.

If no boundary condition is imposed, the non-normalized eigenstate is
simply $\psi(x) = D_\nu(x\sqrt{2m\omega/\hbar})$, where $D_\nu(z)$ is
the parabolic cylinder function (see \ref{sec:relations}), and $\nu =
-c-1/2$.  One can check that
\begin{eqnarray}
\label{eq:Dnu_large1}
D_\nu(z) &\simeq& e^{-z^2/4} z^\nu \biggl[1 - \frac{\nu(\nu-1)}{2z^2} + O(z^{-4})\biggr]  \qquad (z\gg 1), \\
\label{eq:Dnu_large2}
D_\nu(z) &\simeq& e^{-z^2/4} z^\nu \biggl[1 - \frac{\nu(\nu-1)}{2z^2} + O(z^{-4})\biggr] \\
\nonumber
&-& \frac{\sqrt{2\pi}}{\Gamma(-\nu)} e^{\pi i\nu} e^{z^2/4} z^{-\nu-1} \biggl[1 + \frac{(\nu+1)(\nu+2)}{2z^2} + O(z^{-4})\biggr] \quad (z\ll -1) .
\end{eqnarray}
In order to eliminate the unphysical rapid growth of the eigenstate as
$z\to -\infty$, one needs to impose $\nu = n$ with $n = 0,1,2,\ldots$
to remove the last term, from which one retrieves the quantized
energies of the quantum harmonic oscillator: $E_n = \hbar \omega(n +
1/2)$, while the eigenfunctions become expressed through the Hermite
polynomials $H_n(z)$
\begin{equation*}
\psi_n(x) = \frac{1}{\sqrt{2^n n!}} ~ \left(\frac{m\omega}{\pi\hbar}\right)^{1/4} \exp\left(- \frac{m\omega x^2}{2\hbar}\right)
H_n\left(\sqrt{\frac{m\omega}{\hbar}} x\right) ,
\end{equation*}
where the usual normalization prefactor is included, and we used
$D_n(z) = 2^{-n/2} e^{-z^2/4} H_n(z/\sqrt{2})$.  Since imposing no
boundary condition corresponds to barriers at distance $L\to\infty$,
we retrieve the asymptotic behavior $\lambda_n \simeq \frac{k}{\gamma}
n$ or, equivalently, $\alpha_n^2 = \frac{L^2}{D}\lambda_n \simeq
2\kappa n$ as $\kappa\to\infty$.  Note that the prefactor $2\kappa$ is
twice smaller than that of Eq. (\ref{eq:alphan_b}) because the latter
relation accounts only for symmetric eigenfunctions that contribute to
the survival probability.

Imposing Dirichlet boundary condition at $x = \pm L$ corresponds to
setting infinite potential outside the interval $[-L,L]$ (and keeping
the harmonic potential inside).  The eigenvalue problem for a quantum
oscillator in such potential is equivalent to the analysis of the
first exit time distribution in Sec. \ref{sec:spectral_1d}.

%%%%%%%%%%%%%%%%%%%%%%%%%%%%%%%%%%%%%%%%%%%%%%%%%%%%%%%%%%%%%%%%%%%%%%%%%%%%%%%%%%%%%%

\addtocontents{toc}{\protect\setcounter{tocdepth}{1}}

\end{document}